\documentclass[12pt]{iopart}
\usepackage{graphicx}
\usepackage{cite}
\usepackage[colorlinks]{hyperref}
\usepackage{amstext}  
\usepackage{amssymb}

\usepackage[usenames]{color}

\newcommand{\dip}[1]{\times 10^{#1}}
\newcommand{\ket}[1]{\left|#1\right\rangle}

\newcommand{\Ni}{N_{\text{i}}}
\newcommand{\Nf}{N_{\text{f}}}

\begin{document}
\title{Spontanous spin squezing in a rubidium BEC}
\author{Th{\'e}o Laudat$^1$, Vincent Dugrain$^2$, Tommaso Mazzoni$^1$,
  Meng-Zi Huang$^2$, Carlos L. Garrido Alzar$^1$, Alice
  Sinatra$^2$, Peter Rosenbusch$^1$ and Jakob Reichel$^2$}
\address{$^1$ LNE-SYRTE, Observatoire de Paris--Universit\'e PSL, CNRS, 
   Sorbonne Universit\'e, 
   61 Avenue de l'Observatoire, 75014 Paris, France}
\address{$^2$ Laboratoire Kastler Brossel, ENS--Universit\'e PSL, CNRS, 
   Sorbonne Universit\'e, Coll\`ege de France,
   24 rue Lhomond, 75005 Paris, France}
\ead{jakob.reichel@ens.fr}

\date{\today}

\begin{abstract}
  We describe an experiment where spin squeezing occurs spontaneously
  within a standard Ramsey sequence driving a two-component
  Bose-Einstein condensate (BEC) of $^{87}$Rb atoms trapped in an
  elongated magnetic trap. Multiparticle entanglement is generated by
  state-dependent collisional interactions, despite the near-identical
  scattering lengths of the spin states in $^{87}$Rb. In our
  proof-of-principle experiment, we observe a metrological spin
  squeezing that reaches $1.3\pm 0.4\,$dB for $5000$ atoms, with a
  contrast of $90\pm 1$\%. The method may be applied to realize
  spin-squeezed BEC sources for atom interferometry without the need
  for cavities, state-dependent potentials or Feshbach resonances.
\end{abstract}

\section*{Introduction}

Bose-Einstein condensates (BECs) of atoms with more than one spin
state
present rich dynamics in their spin and motional degrees of freedom.
Most experiments so far have focused on either spin or real-space
evolution, carefully avoiding time-dependent evolution in the other
subspace.  One class of experiments prepares each atom in the BEC in a
precisely controlled spin superposition and explores the complex
spatial phase dynamics that deploys due to the spin-dependent
interactions \cite{Kawaguchi12}. In these studies, mainly focused on
the mean field dynamics, the spin state remains unchanged throughout
the evolution. Other experiments, by contrast, use the condensed
sample as a support for spin dynamics, especially to generate
entangled spin states, while spatial dynamics is carefully avoided
\cite{Esteve08,Gross10}. Only a few recent experiments have started to
explore the interplay of spatial and spin dynamics in order to
generate different forms of entanglement in two-component or spinor
BECs \cite{Riedel10,Luecke11,Schmied16}, including spin squeezing.
Spin-squeezed states \cite{Kitagawa93,Wineland94} are the prime
example of highly entangled many-particle states with the potential to
improve atomic clocks and interferometric sensors beyond the standard
quantum limit \cite{Gross12}. This metrological prospect also applies
to BECs which, due to their minimum phase-space spread, are considered
as precious source states for atom interferometry
\cite{Gross12,Abend16,Hardman16} despite their inherent fluctuations,
phase diffusion and losses. Furthermore, the spin squeezing parameter
can be used to quantify the degree of entanglement between the
condensate atoms \cite{Sorensen01}. For all these reasons,
spin-squeezed states of BECs have met with wide interest, and it has
been pointed out early on that such states can naturally arise in
two-component BECs due to different scattering lengths between the
internal states \cite{Sorensen01}. Yet, experiments with two-component
BECs have not produced such states, except when atomic interactions
were enhanced with the help of a Feshbach resonance \cite{Gross10} or
by actively separating the spin components in a state-dependent trap
\cite{Riedel10}. Both methods have led to spectacular results, but
come at the price of a considerably more complex setup. Here we
describe an experiment where spin squeezing occurs spontaneously after
an internal state quench, the dynamics being initiated simply by an
initial $\pi/2$ pulse \cite{Haine14} applied to a rubidium BEC in a
harmonic trap.

\section{Origin of spontaneous spin squeezing}
The basic idea of creating spin squeezing by atomic interaction in a
BEC, as originally envisaged in 2001 \cite{Sorensen01}, is easily
understood in the basis of well-defined atom numbers $\ket{N_1}$ and
$\ket{N_2}$=$\ket{N-N_1}$, where the index refers to the spin state
and $N$ is the total atom number, which we consider fixed for now. On
the $N$-atom Bloch sphere, each state with a given $N_1$ corresponds
to a circle of fixed latitude. If the energy of these states depends
monotonically on $N_1$, a superposition of several $\ket{N_1}$, such
as a coherent state, will not evolve with constant phase speed on the
Bloch sphere, but will be sheared. This leads to spin squeezing due to
the well-known ``one-axis twisting'' Hamiltonian
\cite{Kitagawa93}. More precisely, for a BEC with spin states $i=1,2$
having spatial wavefunctions $\phi_i(\mathbf r)$, and
$S_z=(N_2-N_1)/2$, the spin interaction can be written
\begin{equation}
H_{\text{int}}/\hbar = \chi S_z^2\,.
\label{eq:Hint}
\end{equation}
Neglecting the dependence of the condensate mode on atom number,
$\chi$ can be written simply \cite{Pezze16}\footnote{In general, $\chi$ can be
  expressed as the derivative of the condensate relative phase with
  respect to the relative number of particles \cite{Li09}, and in
  stationary conditions
  $U_{jk}=-\frac{1}{2\hbar}\partial_{N_j} \mu_k$.}.
\begin{equation}
\chi = (U_{11}+U_{22}-2U_{12})/(2\hbar)\qquad \mbox{and}\qquad
U_{jk} = g_{jk}\int dr^3|\phi_j|^2|\phi_k|^2
\end{equation}
with $g_{jk}=4\pi \hbar^2 a_{jk}/m$ and $m$ the mass of the
atom. Significant squeezing develops for times $t$ such that
$\chi t \ge \frac{1}{N}$ \cite{Sinatra12a}. However, when all
scattering lengths are nearly equal,
$a_{11}\approx a_{22}\approx a_{12}$ as in the case of $^{87}$Rb, and
there is full spatial overlap between the components, population
imbalance causes only a small energy change, and $\chi$ becomes so
small that the required $t$ is unrealistically large. This rules out
the straightforward implementation of BEC squeezing in $^{87}$Rb --
the most widely used atom in two-component BEC experiments and in
cold-atom metrology today. If, on the other hand, the overlap of the
components is reduced, then a sizeable nonlinear interaction exists
even for identical scattering lengths
\cite{Riedel10,Maussang10,Haine14}.

Under properly chosen trapping conditions, spatial dynamics of the
spin components will occur spontaneously \cite{Mertes07}, reducing the
overlap and thus creating spin squeezing. Note that the spatial
dynamics is created by the same difference of scattering lengths
which, while too weak to create spin squeezing on its own, can be
strong enough to drive the spatial separation which then causes the
squeezing. $\chi$ dynamically increases during the separation,
generating the squeezing, and then decreases again as the atoms
oscillate back to their initial position.

\begin{figure}[!ht]
\centering
\includegraphics[width=0.8\columnwidth]{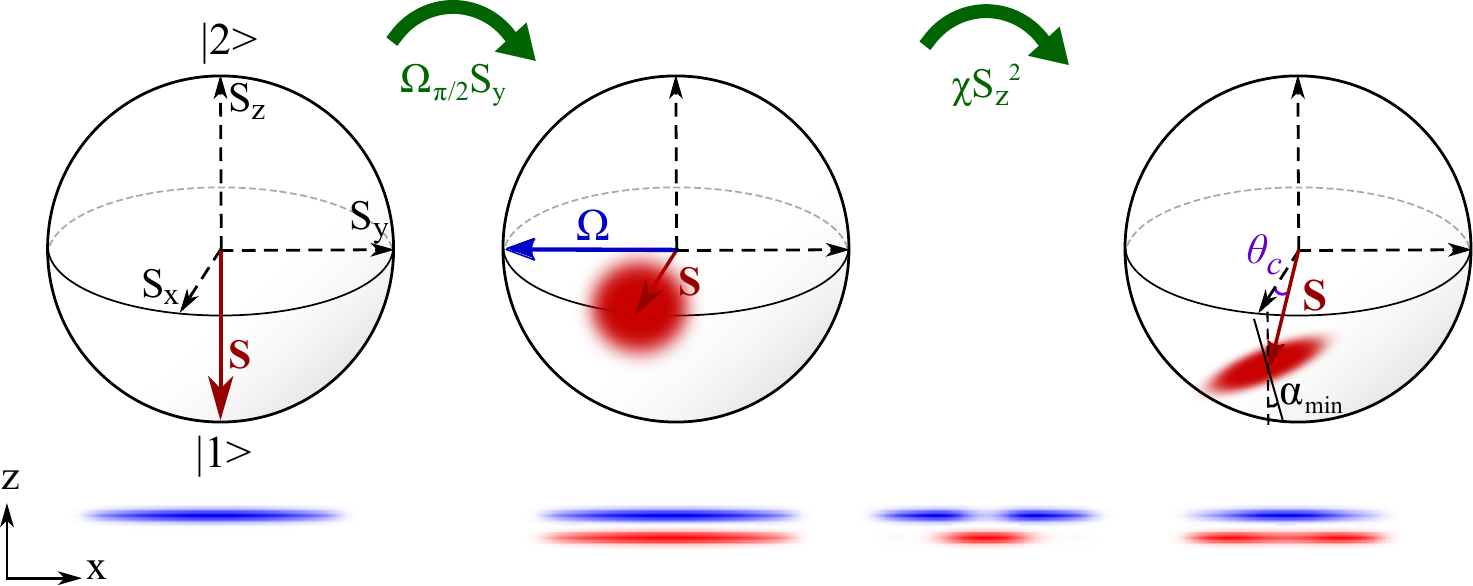}
\caption{Experimental sequence. A first $\pi/2$ pulse of Rabi
  frequency $\Omega$ places the condensate in a coherent
  superposition. This initiates state-dependent spatial dynamics,
  leading to the shearing of the spin noise distribution. Due to
  asymmetric losses, the mean spin is also tilted below the equatorial
  plane by an angle $\theta_c$. A second pulse of variable duration
  is applied in order to rotate the spin distribution before detecting
  the atom numbers $N_1$ and $N_2$. The clouds below the Bloch spheres
  represent the spatial dynamics undergone by the two states $\ket{1}$
  (blue) and $\ket{2}$ (red).}
\label{fig:scheme}
\end{figure}

In this article, we experimentally demonstrate this effect.  In our
experiment, an elongated trap is operated near the ``magic'' bias
field \cite{Harber02,Szmuk15} where trapping frequencies are identical
for two hyperfine ground state sublevels
$\ket{1}\equiv\ket{F=1,m_F=-1}$ and $\ket{2}\equiv\ket{F=2,
  m_F=1}$. The condensate is initially prepared in $\ket{1}$ and
subjected to a $\pi/2$ pulse on the $\ket{1}\leftrightarrow\ket{2}$
transition.  The subsequent free evolution in the cigar-shaped trap
leads to demixing of the two components
\cite{Hall98,Mertes07,Egorov11,Haine14,Nicklas15}, initiating the
squeezing dynamics (Fig.~\ref{fig:scheme}). By applying a second pulse
to close the spin interferometer when the $\ket{1}$ component
oscillates back into overlap with $\ket{2}$, we indeed observe not
only a contrast revival, but also a simultaneous reduction of spin
projection noise, yielding metrological spin squeezing.

\section{Experiment}

The experiment is performed on the Trapped-Atom Clock on a Chip (TACC)
platform, described in detail in
\cite{Lacroute10,Deutsch10,Szmuk15}. In contrast to those references,
here we use a BEC. An atom chip generates the magnetic field gradients
for trapping and also carries the two-photon, radiofrequency (RF) and
microwave (MW) signals for exciting the clock transition. Atoms are
initially trapped in $\ket{1}$ and cooled by forced RF evaporation in
a tight trap.  We continue the RF ramp well into the BEC regime,
obtaining condensates with no discernible thermal fraction and
containing up to $\sim 14000$ atoms. The magnetic potential is then
slowly (600\,ms) transformed into an interrogation trap with
frequencies $\omega_{x,y,z}=2\pi\times(2.7,92,74)\,\mbox{Hz}$ unless
otherwise specified, located $z=350\,\mu$m below the chip surface
($z=0$). The lifetime of the BEC in this trap is about $5\,s$, limited
by collisions with thermal atoms in the single vacuum cell. The atom
number in this trap is controlled with the MOT loading time and the
final frequency of the evaporation ramp.

\begin{figure}
\centering
\includegraphics[width=0.8\columnwidth]{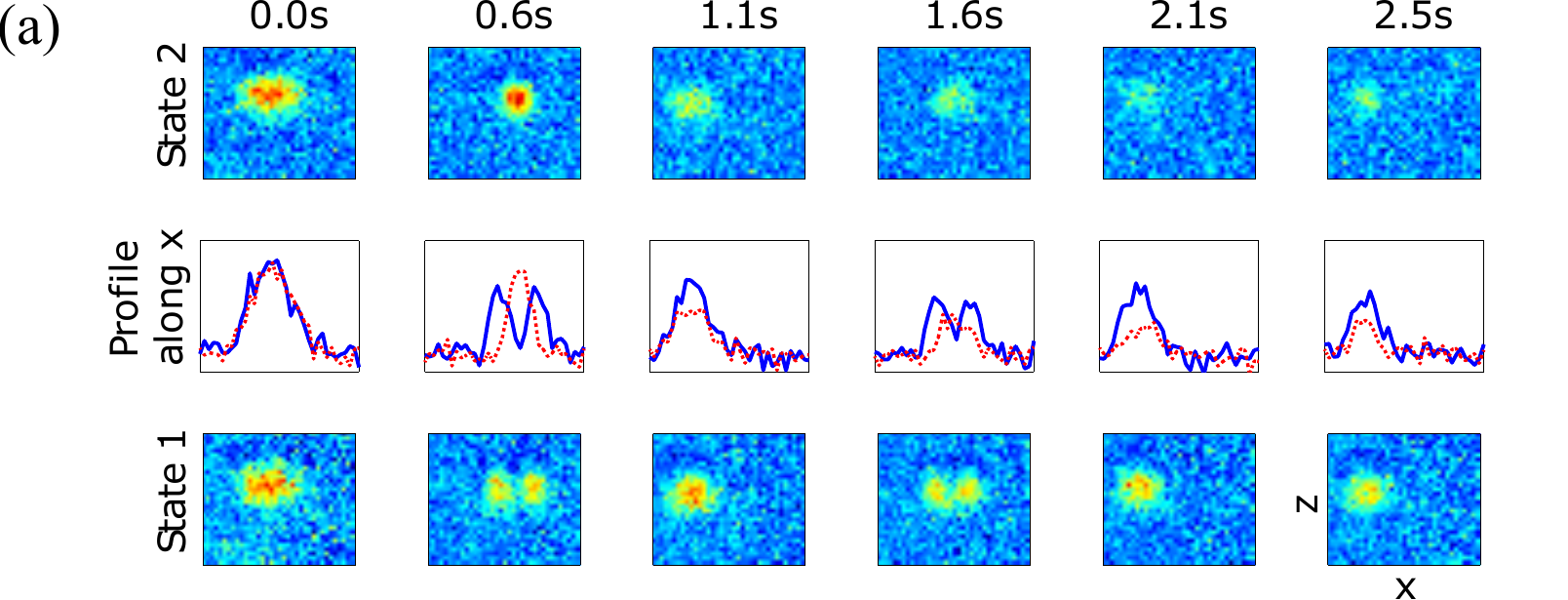}
\includegraphics[width=0.8\columnwidth]{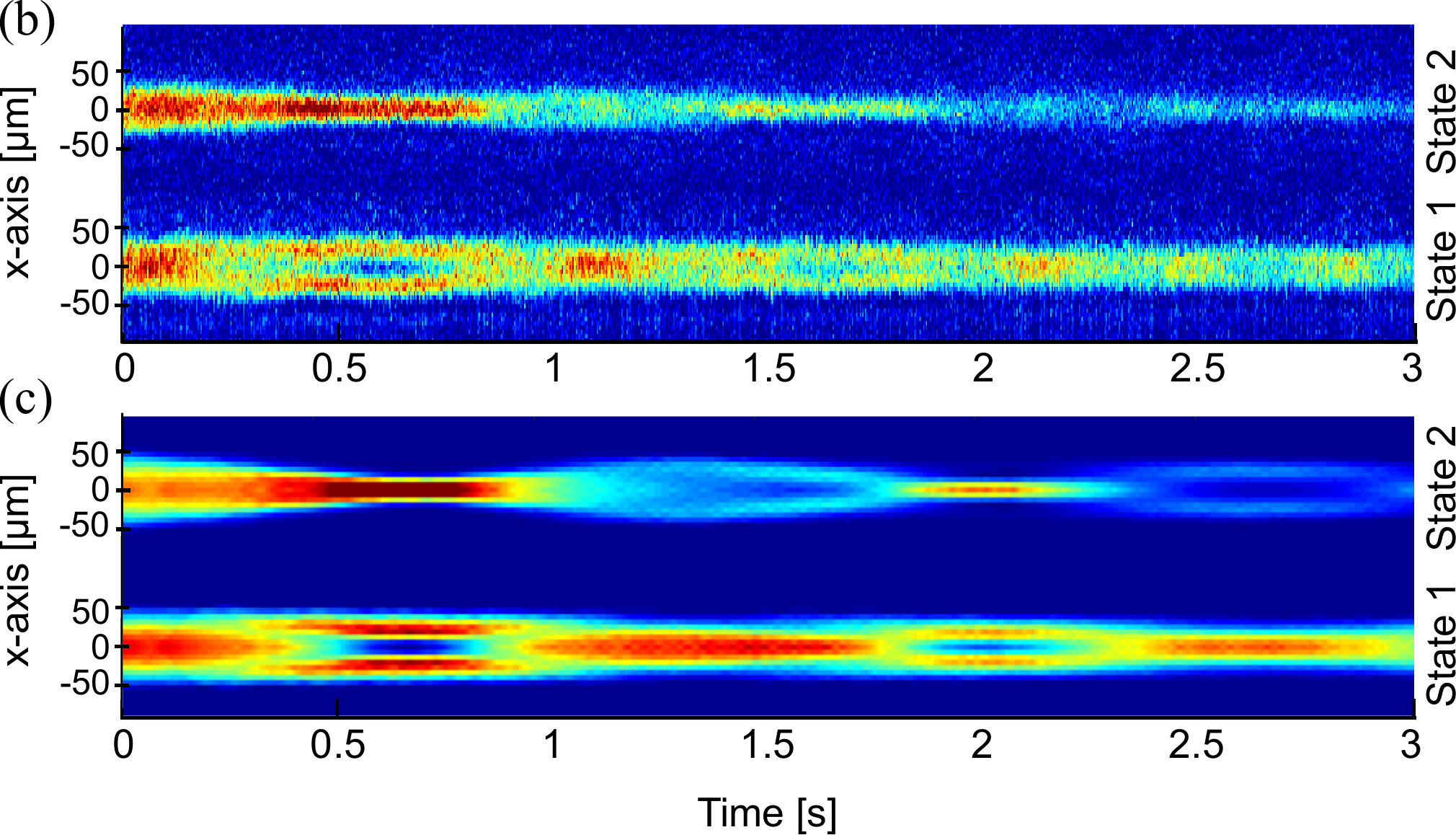}
\caption{Spatial dynamics observed in absorption imaging after a 30\,ms
  time of flight. Note that these images are taken with an auxiliary
  imaging system along the $y$ axis, which has higher noise than the
  one used for the squeezing measurements below. 
  For these measurements, a BEC of $10^{4}$ atoms is
  produced in state $\ket{1}$ in a trap with trap frequencies
  $\omega_{x,y,z}=2\pi\times(2.7,92,74)\,\mbox{Hz}$. A resonant $\pi/2$ pulse
  prepares an equal superposition of $\ket{1}$ and $\ket{2}$ and the
  cloud dynamics are monitored in time. 
  (a) Individual images taken after the evolution times indicated in
  the figure. (b) Many such images integrated along $z$ and assembled to
  show the spatial dynamics along $x$. (A common-mode sloshing that was
  present in this experiment has been subtracted.) (c) 3D coupled
  Gross-Pitaevskii numerical simulation for the atom numbers and trap
  frequencies of the experiment.}
\label{fig:TypicalImages}
\end{figure}

We use the $\ket{1} \leftrightarrow \ket{2}$ clock transition, which
enables first-order cancellation of spatial inhomogeneity of the
transition frequency in a magnetic trap
\cite{Harber02,Treutlein04}. The transition is driven by a two-photon,
RF and MW pulse \cite{Szmuk15} with Rabi frequency
$\Omega = 2\pi\times 3.6\,$Hz.  The MW signal at 6.8\,GHz is generated
by a custom-built synthesizer \cite{Ramirez10}, while the RF photon of
$\approx 2\,$MHz comes from a commercial direct-digital
synthesizer. Both are referenced to SYRTE's active hydrogen maser
\cite{Szmuk15}. After preparing a BEC in the interrogation trap, the
sequence always starts by applying a resonant $\pi/2$ pulse to create
the superposition $1/2^{N/2}(\ket{1}+\ket{2})^{\otimes N}$ (see
\ref{sec:appendixExp}). Due to the slight difference in scattering
lengths, the initial density distribution no longer corresponds to a
stationary state, and the two components start to oscillate
\cite{Hall98,Mertes07,Papp08,Tojo10,Egorov13,Nicklas15}. To reveal the
resulting spatial dynamics, we have imaged both states using an
auxiliary imaging system on the $y$ axis, so that the slow $x$ axis
is visible. Images are taken at variable times after the pulse. A
typical result is shown in Fig.~\ref{fig:TypicalImages}.  The
$\ket{1}$ component splits into two parts which oscillate along the
weak axis, while the $\ket{2}$ component does not separate but
undergoes a breathing-type oscillation in the center between the
$\ket{1}$ component's two lobes. After a period of $1.2\,$s, the
$\ket{1}$ component has come back into superposition with $\ket{2}$
and another oscillation begins. For longer times, a third oscillation
is barely visible. A 3D numerical simulation using coupled
Gross-Pitaevskii equations (GPEs) reproduces the features of the
observed oscillation (Fig.~\ref{fig:TypicalImages}(b)) reasonably
well, however, the calculated and measured oscillation frequencies
differ by 20\,\%. One possible reason for this difference could be a
residual thermal cloud too weak to be visible on the camera images.

\begin{figure}
\begin{center}
\includegraphics[width=0.45\columnwidth]{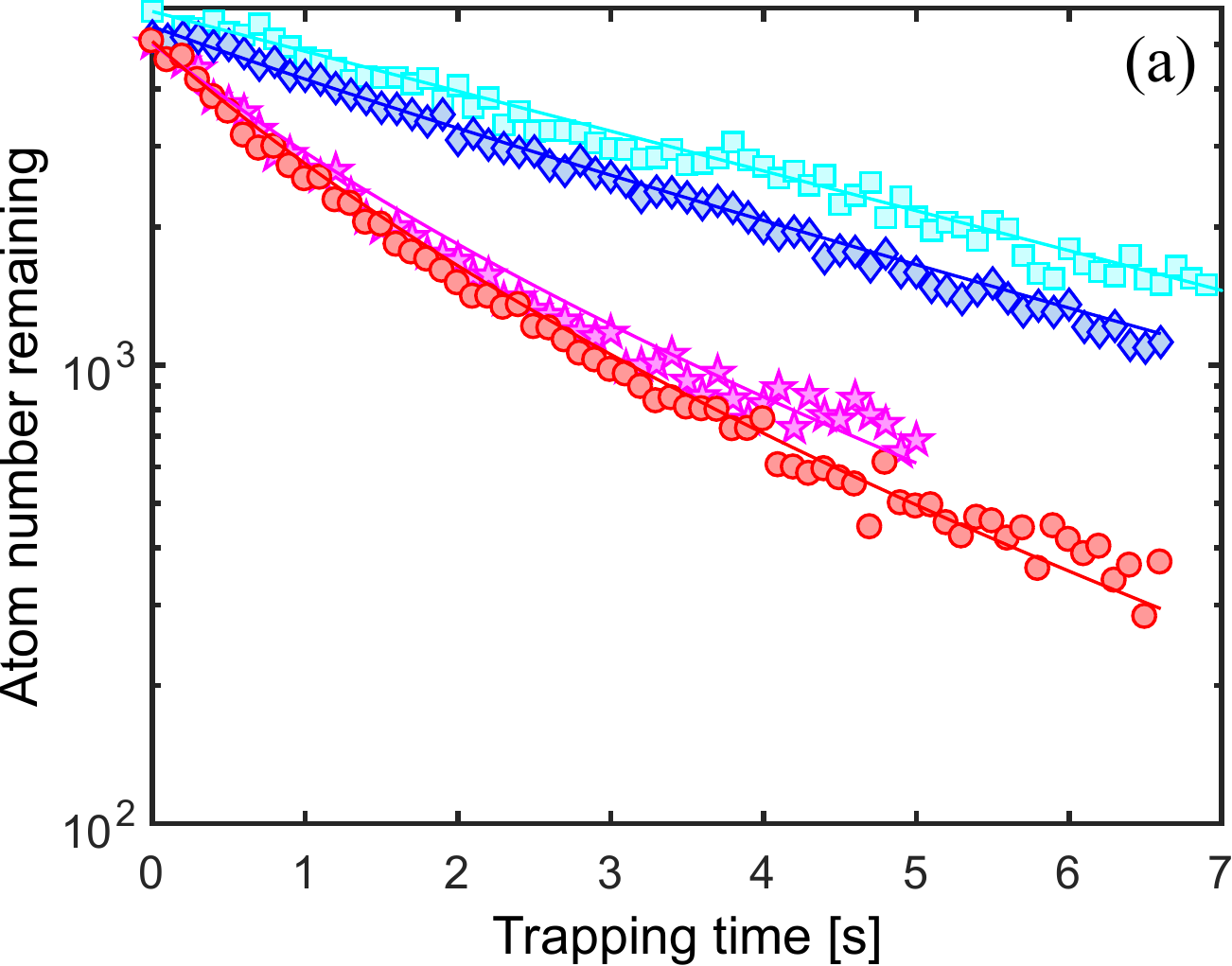}
\includegraphics[width=0.45\columnwidth]{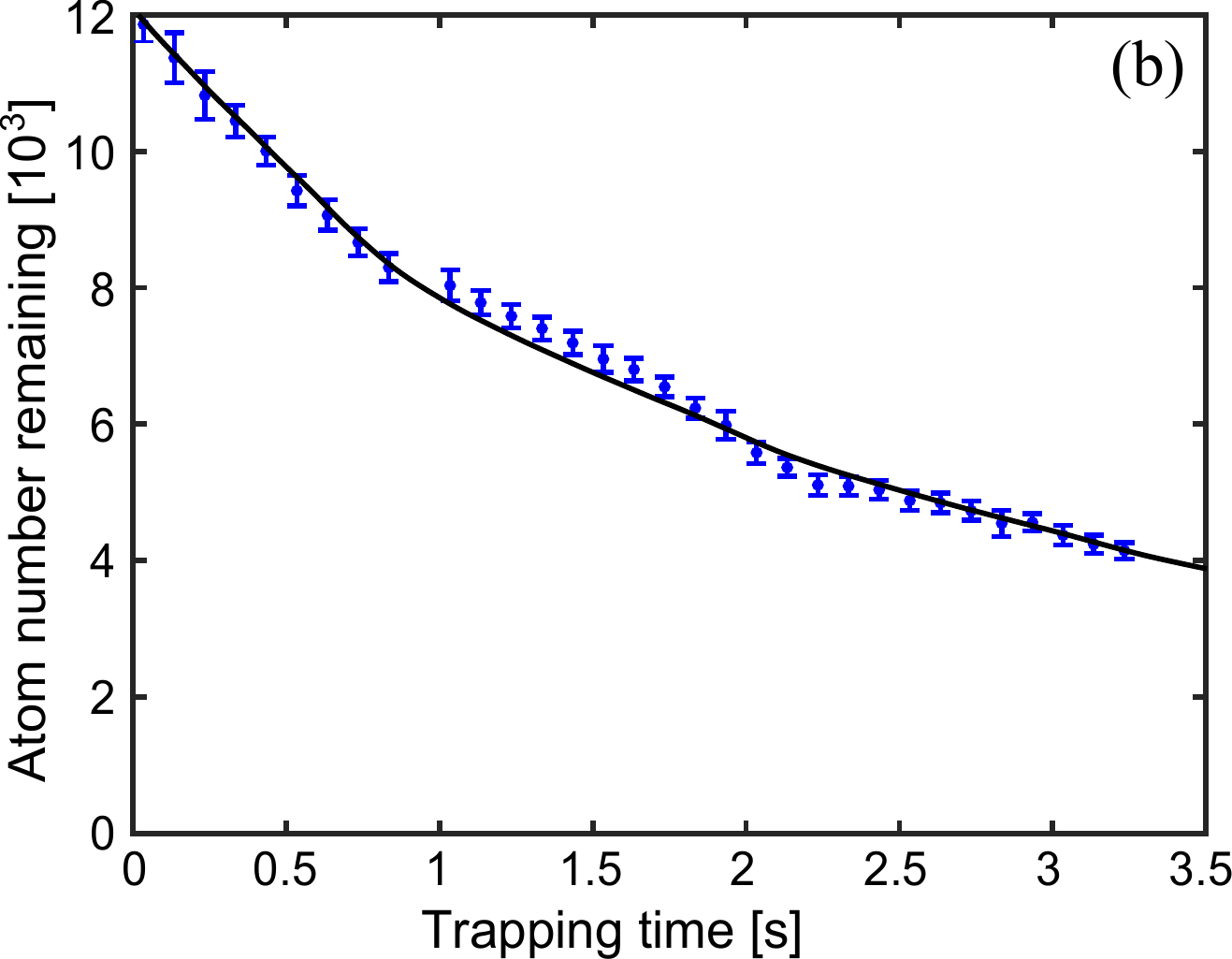}
\end{center}
\caption{Atom number as a function of trapping time in the
  interrogation trap for condensates with all atoms in $\ket{1}$ (cyan
  squares), all atoms in $\ket{2}$ (magenta stars) and with atoms in
  an equal superposition (blue diamonds: $N_1$, red circles:
  $N_2$). The fit to the $\ket{1}$ data is a simple exponential,
  yielding the $5$\,s background-limited lifetime. The other lines are
  not fits, but predictions without adjustable parameters, using
  published values \cite{Egorov13} for the two-body decay rates,
  $\gamma_{22} = 8.1(3)\times10^{-14} \mbox{cm}^3/\mbox{s}$ and
  $\gamma_{12} = 1.51(18)\times 10^{-14} \mbox{cm}^3/\mbox{s}$, and
  our experimentally determined densities. The latter are obtained
  from the measured atom numbers and the trap frequencies, assuming a
  BEC in the dimensional crossover regime. (b) Total atom number
  $N_1+N_2$ as a function of time, starting from an initial atom
  number $N\sim 1.2\dip{4}$ in equal superposition of $\ket{1}$ and
  $\ket{2}$. The blue points are experimental data, while the solid
  line has been obtained by numerical integration of coupled
  GPEs.}
\label{fig:losses}
\end{figure}

While these images are instructive for observing the spatial dynamics,
measurements of the atom numbers are better performed by imaging along
the $x$ axis, where the cloud covers fewer pixels. We use saturated
absorption imaging \cite{Reinaudi07} and state-selective release from
the trap so that both states can be detected in the same image with a
back-illuminated deep depletion CCD with high quantum efficiency. The
imaging system is carefully calibrated for absolute accuracy as
described in \ref{sec:appendixExp}.

Density-dependent atom losses are an important limiting factor in BEC
spin squeezing \cite{Gross10,Li08}.  While the background-limited
lifetime of state $\ket{1}$ is much longer than the oscillation
period, state $\ket{2}$ has additional loss channels which reduce its
lifetime. To measure the relevant loss parameters, we prepare a BEC in
$\ket{1}$, $\ket{2}$ or an equal superposition of both states, and
measure the remaining populations in the interrogation trap after
different trapping times. The results are displayed on
Fig.~\ref{fig:losses} (points). An exponential fit to the $\ket{1}$
data yields a $5$\,s background-limited lifetime. The other curves in
Fig.~\ref{fig:losses}(a) are not fits but predictions without
adjustable parameters as described in the caption. They reproduce the
data well, as does the simulation using coupled GPEs
(Fig.~\ref{fig:losses}(b)).  In our experiments, number densities in
the $\ket{2}$ state range from $1\dip{12}$ to
$8\dip{12}\,\mbox{cm}^{-3}$, corresponding to two-body loss limited
lifetimes ranging from $12\,$s to as short as $1.5\,$s,
respectively. In the following, $\Ni$ refers to the initial atom
number and $\Nf$ to the atom number measured at the end of the
experimental sequence.

\section{Oscillation of the Ramsey contrast}
\label{sec:contrast}

\begin{figure}
\begin{center}
\includegraphics[width=0.45\columnwidth]{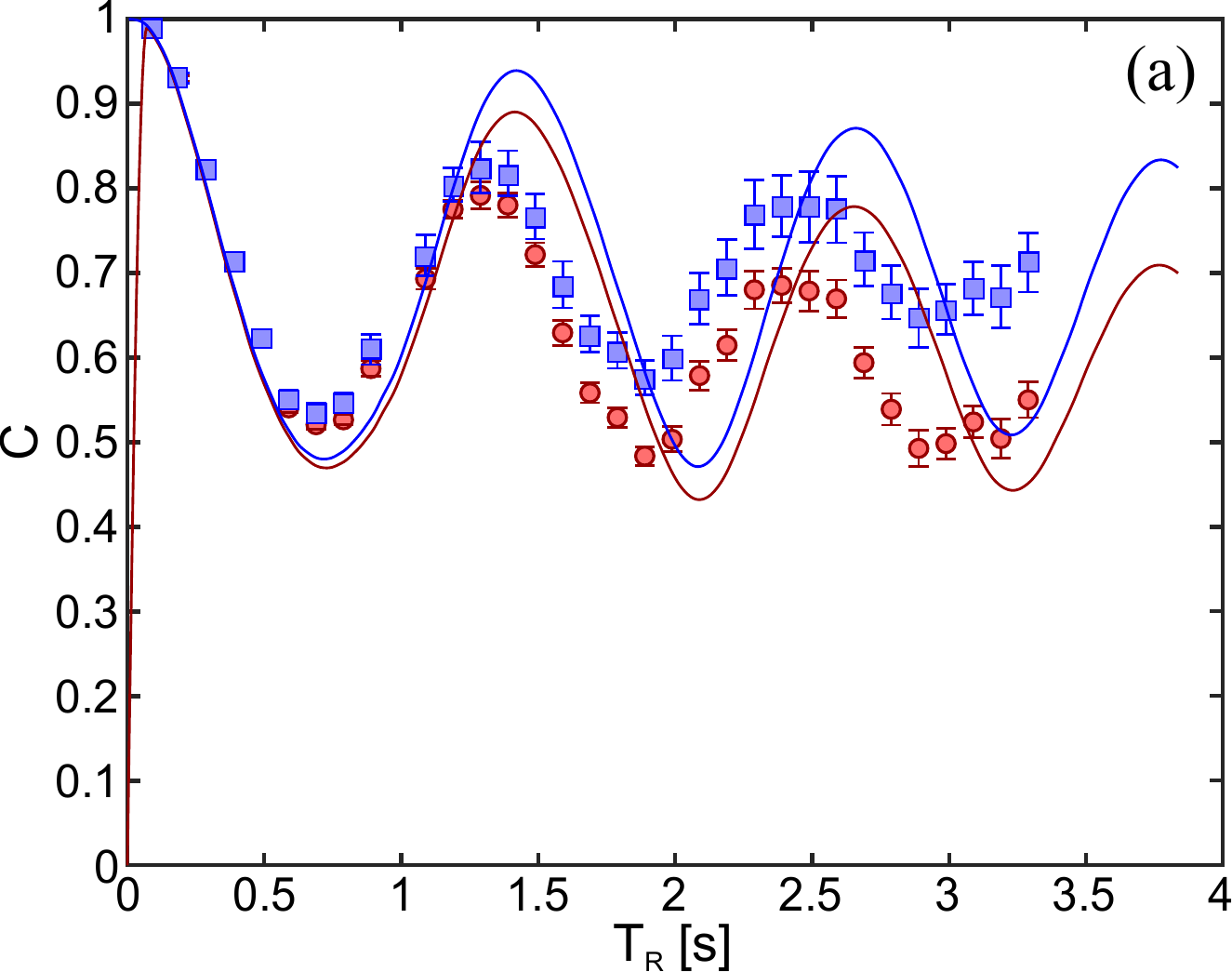}
\includegraphics[width=0.45\columnwidth]{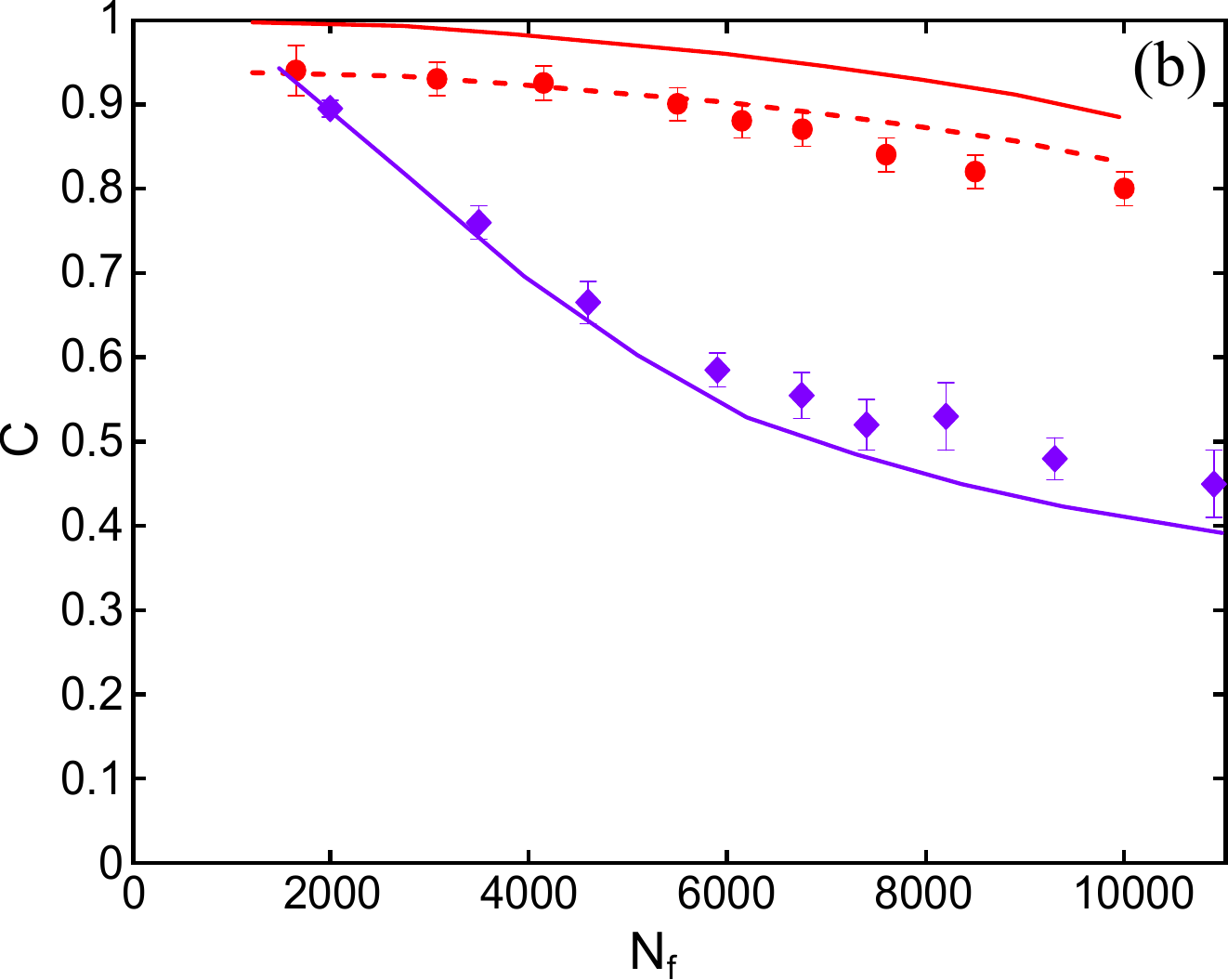}
\end{center}
\caption{Contrast of a Ramsey measurement consisting of two $\pi/2$
  pulses separated in time by $T_R$, performed in the interrogation
  trap with frequencies
  $\omega_{x,y,z}=2\pi\times (2.7,92,74)\,\mbox{Hz}$. For each $T_R$,
  the frequency of the two pulses is scanned around resonance to
  measure the fringe contrast $C$, defined by a fit according to
  $P_2=\frac{1}{2}(1+C\text{cos}(2\pi\Delta\nu T_R+ \varphi_{lo}))$,
  where $P_2=N_2/\Nf$, $T_R$ is the Ramsey time and corresponds to the
  time during which the interferometer is sensitive to phase
  variations, and $\Delta\nu$ is the detuning from atomic
  resonance. (a) Measured contrast (red circles), corrected contrast
  (blue squares) and corresponding predictions of the coupled GPE
  simulation (solid lines), as a function of Ramsey time for
  $\Ni=1.2\times 10^4$. (b) Corrected contrast at the revival time
  (red circles) and half the revival time (purple diamonds) as a
  function of $\Nf$. The corresponding solid lines are results of the
  GPE simulations. The dashed line shows the simulation result
  multiplied with a decoherence term (exponential decay) which
  accounts for decoherence sources not contained in the
  simulation. Its value $\gamma_d=0.5s^{-1}$ is chosen to match the
  maximum experimental contrast.}
\label{fig:contrast}
\end{figure}

Because the oscillation puts the atoms into motion and changes the
spatial overlap of the components, it also manifests itself in the
contrast of the Ramsey fringes when a second $\pi/2$ pulse is added
after a time $T_R$.  Fig.~\ref{fig:contrast}(a) shows the evolution of
this contrast (red circles) as a function of $T_R$ for
$\Ni = 1.2\times 10^4$. For this atom number, the initial contrast of
98\% drops to about 50\% around 600\,ms, and then shows a revival at
$T_R=1.2\,$s , which reflects the spatial overlap between the two
modes at this time. Indeed, we find that the contrast revival time
coincides with the spatial oscillation period observed by absorption
imaging. This period depends on trap frequencies and atom number
\cite{Egorov13}.

Due to the state-dependent losses, the population imbalance is
time-dependent. Right after the initial $\pi/2$ pulse, the polar angle
of the Bloch vector is $\theta=\pi/2$, but then slowly evolves to a
value $\theta=\pi/2+\theta_c$ at time $T_R$ (Fig.~\ref{fig:scheme}).
To obtain maximum contrast (and thus, maximum phase sensitivity in the
final measurement), $\theta_c$ must be taken into account.  In an
actual atomic clock or interferometer, this can be achieved by
inserting a correction pulse with a well-defined phase before the
second $\pi/2$ pulse to remove the known mean value $\bar{\theta}_c$.
The contrast that one would get by applying such a correction pulse
can also be derived numerically using simple geometric considerations
(\ref{sec:appendixExp}). We have used both, correction pulses and
numerical contrast correction, and find that the results are
consistent. Unless otherwise indicated, the results below use the
numerical correction method. Note that only the mean value of
$\theta_c$ can be removed or corrected, while the noise introduced by the
statistical nature of the losses remains and contributes to the final
noise budget \cite{Gross10,Li08}. We will come back to this point
below.

The corrected contrast, represented by blue squares in
Fig.~\ref{fig:contrast}(a), shows a first revival of 82\% for
$\Ni=1.2\times 10^4$. The precise
value of the contrast, and thus the spatial dynamics, also depends on
atom number, as shown in Fig.~\ref{fig:contrast}(b). Lower atom
numbers result in higher contrast revivals.

The numerical simulations described above qualitatively reproduce the
observed time evolution and provide some additional insight (solid
lines in Fig.~\ref{fig:contrast}). The contrast minimum occurring at
half the revival time decays faster than the contrast maximum at
the revival time. Its simulated
value is in good agreement with the experiment, confirming that the
decay is caused by a stronger spatial separation for higher atom
numbers. Both the demixing period and the contrast at revival time are
overestimated in the simulation, even though the population decay is
well reproduced (cf.~Fig.~\ref{fig:losses}(b)). The lower revival
contrast suggests experimental sources of decoherence that are not
contained in the simulation.  Indeed, multiplying with
a decoherence term with $\gamma_d=0.5s^{-1}$ brings the simulated
contrast into agreement with the measured values (dashed line in
Fig.~\ref{fig:contrast}(b)). For the revival period, the source of the
deviation is less obvious, one possible candidate being a dilute
thermal cloud as mentioned above.

Note that the density-dependent frequency shift \cite{Harber02},
combined with the spatial dynamics and atom losses, leads to a time
dependence of the atomic transition frequency $\nu_{12}$: the
resonance frequency of the pulse applied in the beginning of the
sequence is slightly higher than that at the revival time. Although
the shift is small (on the order of $-10^{-4}\,$Hz per atom for our
trap), it is easily detected in a metrology setup like ours and
needs to be taken into account in the squeezing measurements, as
detailed below. In particular, for every atom number and Ramsey time,
we use the adequate effective resonance frequency, which is determined
in a separate measurement (see~\ref{sec:appendixExp}).

\section{Spin noise measurements}

\begin{figure}[tb!]
\centering
\includegraphics[width=\columnwidth]{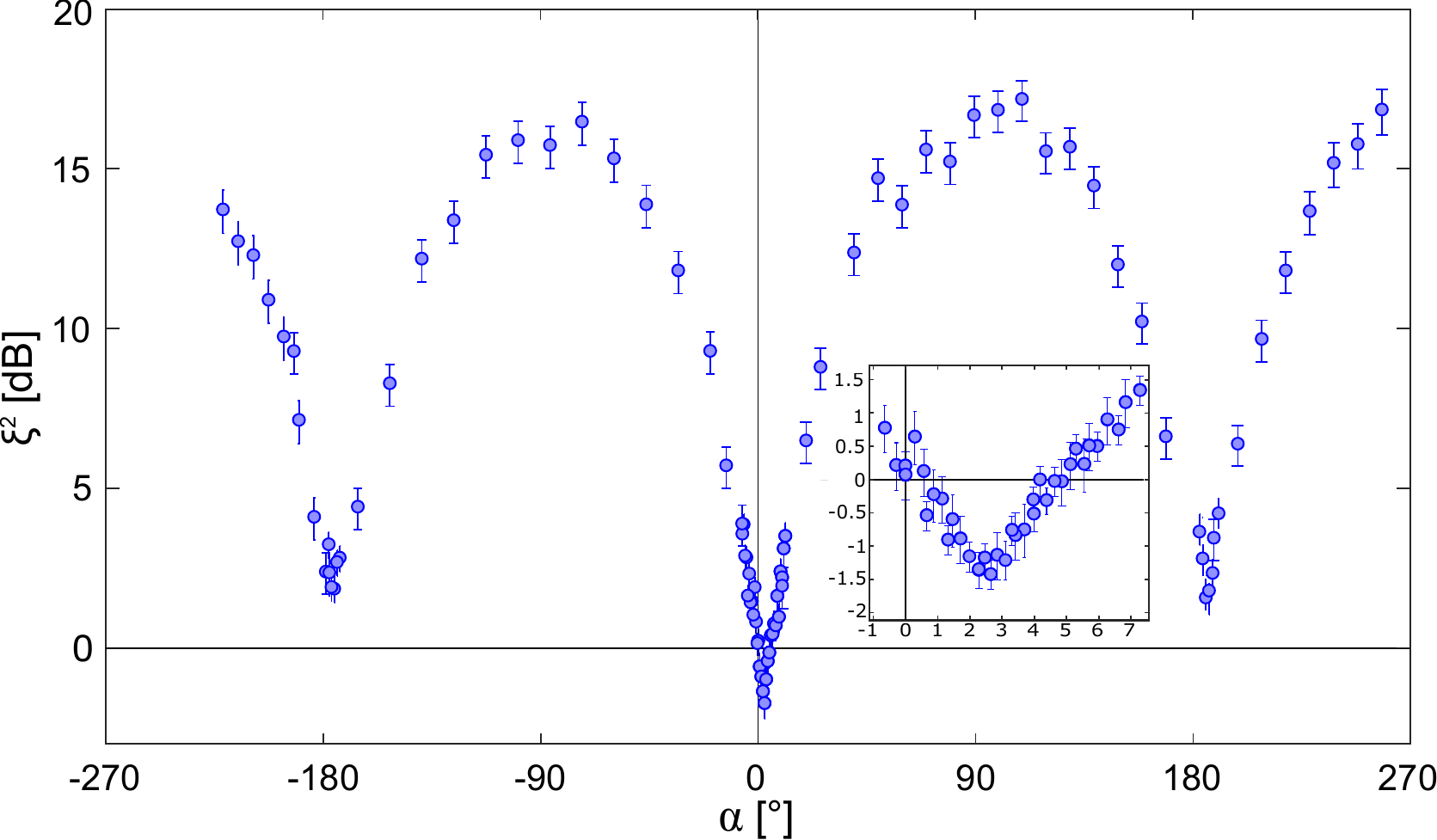}
\caption{Measured spin noise for a final atom number $\Nf\approx5000$
  and trap frequencies
  $\omega_{x,y,z}=2\pi\times (2.7,92,74)\,\mbox{Hz}$. The squeezing
  factor $\xi^2=\Delta_n S_z^2 = 4\Delta S_z^2/C^2\overline{\Nf}$ is
  shown as a function of the tomography angle $\alpha$, with error bars
  corresponding to a 68$\%$ confidence interval. The inset is a zoom
  of the main plot in the region where $\xi^2$ reaches its minimum.}
\label{fig:squeezing_result}
\end{figure}

We use spin noise tomography to characterize the spin distribution
that is generated in the dynamically evolving two-component BEC, . As
before, a BEC with a precisely controlled atom number is produced in
$\ket{1}$ and we apply a first near-resonant $\pi/2$ pulse which puts
each atom into a coherent superposition between the two clock
states. The BEC then evolves freely in the trap during a time $T_R$
which we adjust to coincide exactly with the contrast revival time
measured above. During the free evolution, the spin distribution
undergoes the nonlinear collisional interaction enhanced by the
spatial separation of the two components. At the time $T_R$, a second
pulse (``analysis pulse'') is used to rotate the spin distribution
about its center, which has been determined separately (see
Fig.~\ref{fig:scheme} and \ref{sec:appendixExp}). By changing the
duration of the analysis pulse, the rotation angle $\alpha$
(``tomography angle'') can be varied. After this rotation, the trap is
switched off and the atom numbers $N_1$ and $N_2$ are measured.  For
each $\alpha$, the whole preparation and analysis sequence is repeated
a large number of times (typically 300 repetitions) and the normalized
population difference $S_z^n = \frac{1}{2}\frac{N_2-N_1}{N_1+N_2}$ is
determined. The number squeezing
$\mathcal{V}^2 = \frac{4\Delta
  S_z^2}{\overline{\Nf}\text{cos}(\overline{\theta_c})^2}$ and the
squeezing factor
$\xi^2=\frac{4\Delta
  S_z^2}{\overline{\Nf}C^2\text{cos}(\overline{\theta_c})^2}$
\cite{Wineland94} are then derived in order to quantify the spin noise
reduction and the metrologically useful spin squeezing respectively,
$\Delta S_z^2$ being the variance of the spin in the y-z plane
(Fig.~\ref{fig:scheme}). The contrast is determined separately using
the procedure detailed on
Fig.~\ref{fig:contrast}. Fig.~\ref{fig:squeezing_result} shows the
result for a final atom number $\Nf\approx5000$ and trap frequencies
$\omega_{x,y,z}=2\pi\times (2.7,92,74)\,\mbox{Hz}$. As expected, the
measured noise corresponds to a slightly tilted, ellipse-shaped
distribution. The minimum squeezing factor occurs for an angle
$\alpha=2.5^\circ$ and reaches $\xi^2 = -1.3 \pm 0.4$ dB with a
contrast of 90$\pm$1$\%$ for this parameter set. It corresponds to
atom number fluctuations of $\pm 32$ atoms for each component.

\begin{figure}[tb!]
\centering
\includegraphics[width=0.47\columnwidth]{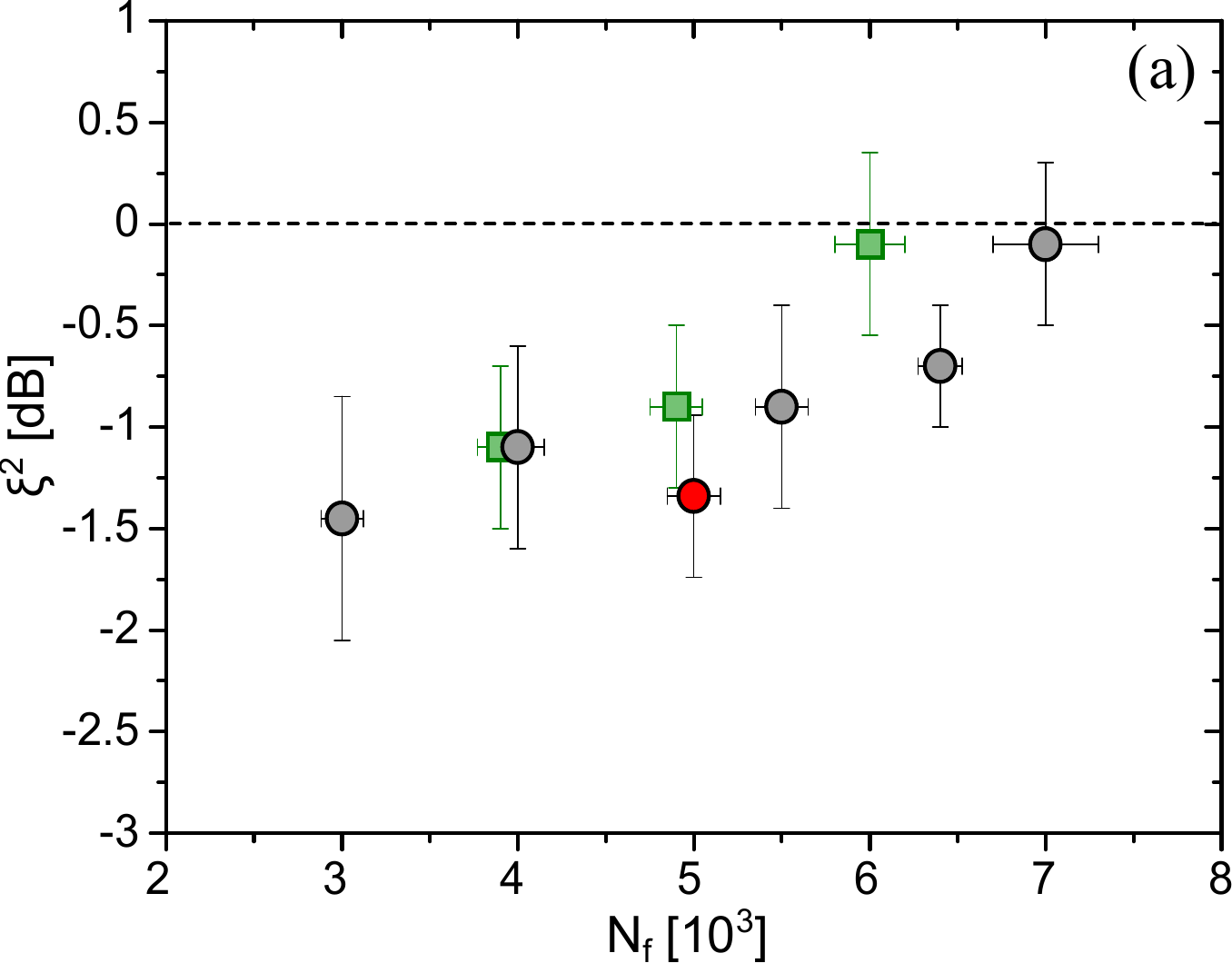}
\includegraphics[width=0.47\columnwidth]{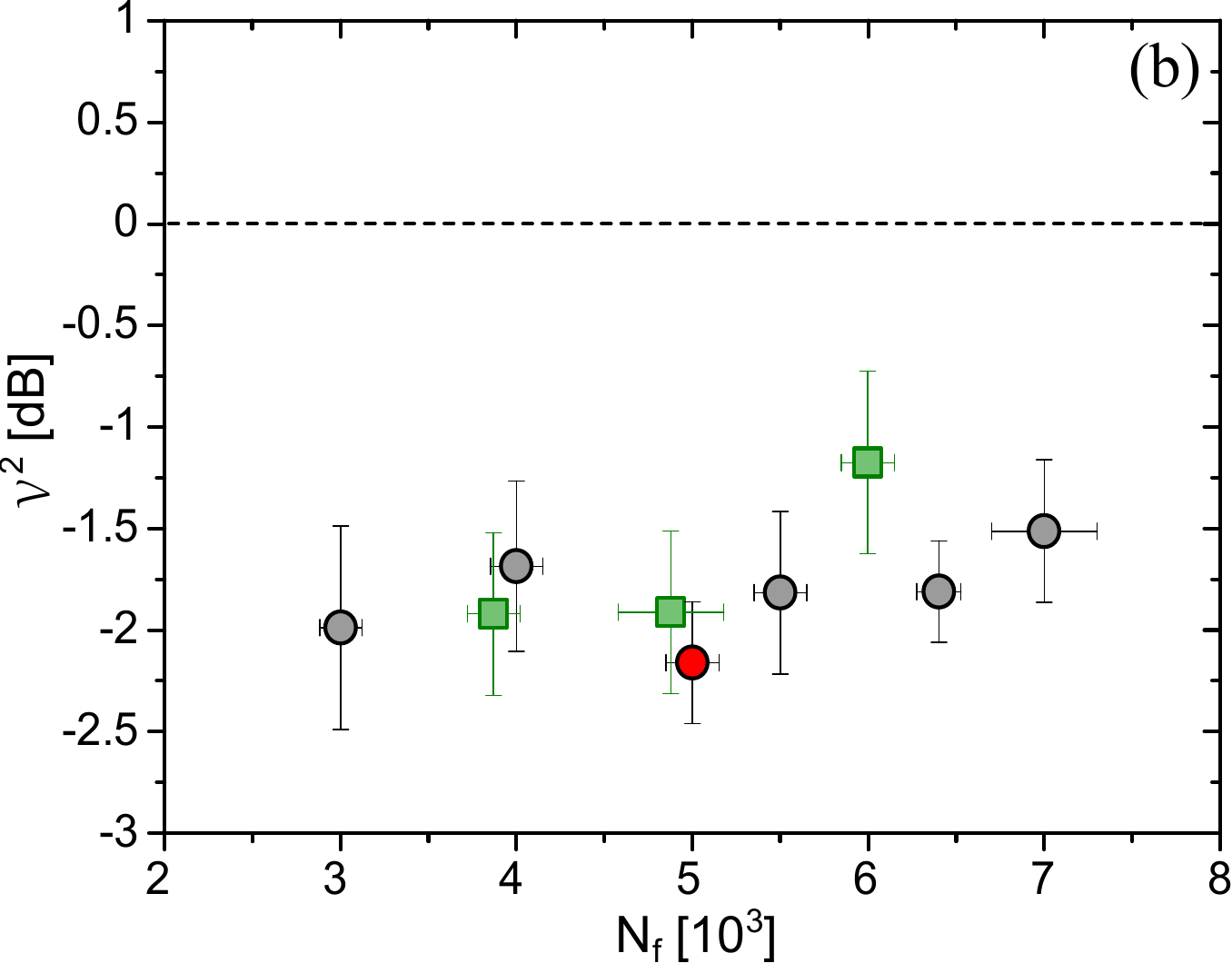}
\caption{Squeezing factor (a) and number squeezing (b) as a function
  of the detected atom number for two different traps. Black circles:
  $\omega_{x,y,z}=2\pi\times (2.7,92,74)\,\mbox{Hz}$, green squares:
  $\omega_{x,y,z}=2\pi\times (4.4,128,113)\,\mbox{Hz}$.  The red point
  corresponds to the data displayed on
  Fig.~\ref{fig:squeezing_result}. The result for the stronger trap
  are generally worse in spite of the faster dynamics
  ($T_R=0.7$s). While the metrological squeezing factor deteriorates
  with increasing atom number, no clear tendency is visible in the number
  squeezing. This indicates that the squeezing factor is
  mostly limited by the contrast reduction occurring when increasing
  the atom number, as shown in Fig.~\ref{fig:contrast} (b).}
\label{fig:squeezingVsNat}
\end{figure}

We have repeated these measurements for different atom numbers up to
the maximum BEC atom number accessible in our experiment, and for a
second, stronger trap with frequencies
$\omega_{x,y,z}=2\pi\times (4.4,128,113)\,\mbox{Hz}$. The results are
shown in Fig.~\ref{fig:squeezingVsNat}. The squeezing factor $\xi$
deteriorates for $\Nf>5000$, but seems to saturate for smaller
$\Nf$. Interestingly, the number fluctuations
(Fig.~\ref{fig:squeezingVsNat}(b)) do not show these tendencies, but
maintain a constant level within the error bars. The deterioration of
$\xi$ is mostly due to the reduced contrast at high atom numbers
(Fig.~\ref{fig:contrast}(b)).  Both effects will be discussed in the
next section.

\section{Limiting factors}
The squeezing factor $\xi$ observed at the revival time results from
the competition between the twisting interaction (eq.~\ref{eq:Hint})
and the state-dependent losses and non-perfect spatial revival
dynamics: the latter two introduce new fluctuations and reduce
contrast.  The two main parameters that can be experimentally
controlled are the atom number $N$ and the trap frequencies
$\omega_{x,y,z}$. Higher atom numbers and higher trap frequencies
increase the condensate density, which accelerates the squeezing
dynamics by increasing $\chi$, but also accelerates the two-body
losses. In the case of a homogeneous system or in separated harmonic
traps for the two components, the two effects cancel
\cite{Li08,Sinatra12a}, so that one does not expect a density
dependence of the squeezing factor $\xi$. In our case, the spatial
dynamics lead to a significantly more complicated situation. $\chi$ as
well as the contrast depend on the spatial overlap of the spin
components, which is time-dependent (cf.~Fig.~\ref{fig:contrast}) with
an evolution that depends on $N$ as well as on $\omega_{x,y,z}$. The
rather high value of the contrast at $T_R/2$ (Fig.~\ref{fig:contrast})
indicates that the component separation is not complete. For complete
separation and our range of atom numbers, it is known that $\chi$
would be large enough to reduce number fluctuations by several orders
of magnitude in a time much shorter than our revival time (see
\ref{sec:appendixSimul}), even when losses are taken into account.  In
the absence of losses, these high values could still be reached with
incomplete separation, at the expense of a longer squeezing
time. Thus, in our situation, the state-dependent losses
(Fig.~\ref{fig:losses}) clearly have a major effect on the final
result. In an attempt to obtain more quantitative predictions, we have
performed beyond-GPE simulations, described in the next section.

Apart from these fundamental contributions, technical noise such as
phase noise can limit the measurement of the noise reduction induced
by the squeezing process. In order to evaluate our system in terms of
technical instabilities, a standard clock measurement, similar to the
one conducted in \cite{Szmuk15}, has been performed using the same
experimental condition as for Fig.~\ref{fig:squeezing_result}. This
measurement yielded a fractional frequency stability of
$9.7\times 10^{-12}\tau^{-1/2}$. Several noise sources have been
investigated to explain this stability, and atom loss has been
identified as the major contribution to the stability budget
($8.47\times 10^{-12}\tau^{-1/2}$). This is due to the fact that, for
each shot, we only have access to the final populations. We therefore
do not precisely know how many atoms have been lost during the
sequence, nor when they were lost. For instance, if an atom is lost at
the beginning of the Ramsey time, it will not contribute to the
collisional shift, whereas if it is lost right before the second
interrogation pulse, it was partly responsible for this frequency
shift, but will not be detected. This leads to a noise on $S_z$ that
we cannot correct. This noise also impacts the squeezing measurement,
contributing on the order of 9\% of the quantum projection
noise. Subtracting this noise would bring $\xi^2$ from $-1.3\,$dB to
$\xi^2 \approx -2\,$dB: its contribution is non-neglegible, but it
does not limit the order of magnitude of the observed squeezing.

\section{Simulations beyond GPE}
While the spatial dynamics is well described by the coupled
Gross-Pitaevskii simulations mentioned above, the quantum spin
dynamics generating the spin squeezing cannot be captured by such a
mean-field approach. Furthermore, as we have seen, the asymmetric
losses significantly affect the state of the system over the
relatively long times needed for the spontaneous spin squeezing to
occur. In order to take into account all of these features in a
consistent way, we performed simulations using a Wigner method
inspired by \cite{Opanchuk12}.  To limit the drawbacks of the
truncated Wigner method \cite{Sinatra02} -- which are related to the
fact that the added quantum noise in each mode efficiently thermalizes
in 3D, introducing spurious effects -- we implemented a ``minimal
version'' of the Wigner method, where we project the quantum noise on
the condensate mode for each component.

\subsection{Description of the projected Wigner method}
The implementation of the method consists in $(i)$ generating
classical fields $\psi_1(\mathbf{r},0^+)$ and $\psi_2(\mathbf{r},0^+)$
normalized to the atom number in each component, that sample the
initial probability distribution after the pulse, and $(ii)$ evolving
them with stochastic equations. Besides the usual Hamiltonian terms,
these equations involve a damping term due to non-linear losses and
the associated noise. The results for the observables are then
obtained by averaging over many stochastic realizations.

\subsubsection{Initial state with partition noise}
At $t=0^-$, before the mixing pulse, all $N$ particles are in the
internal state $\ket{1}$ where a condensate of wave function
$\phi_1(\mathbf{r},0^-)$ is present. We approximate the field for
$\ket{1}$ by $\psi_1(\mathbf{r},0^-)=\sqrt{N}\phi_1(\mathbf{r},0^-)$.
The field for $\ket{2}$ is in vacuum, that is, it is filled with
quantum noise in each mode. In contrast to what is usually done in the
Wigner method, we project the vacuum fluctuations of field 2 on the
condensate mode $\phi_1(\mathbf{r},0^-)$ that will be macroscopically
populated after the pulse, and we keep only this contribution. We then
obtain after the mixing pulse
\begin{eqnarray}
\psi_1(\mathbf{r},0^+)&=&\frac{1}{\sqrt{2}} \left[\psi_1(\mathbf{r},0^-)-\phi_1(\mathbf{r},0^-)b  \right]\\
\psi_2(\mathbf{r},0^+)&=&\frac{1}{\sqrt{2}}  \left[\psi_1(\mathbf{r},0^-)+\phi_1(\mathbf{r},0^-)b  \right]
\end{eqnarray}
where $b$ is a stochastic complex Gaussian variable with $\langle b^\ast b \rangle=\frac{1}{2}$.
As $N_j(t)=\int \mathbf{dr} |\psi_j(\mathbf{r},t)|^2$,  one has $\langle N_1-N_2\rangle(0^+)=0$ and $\Delta^2(N_1-N_2)(0^+)=N$.

\subsubsection{Time evolution}
Starting from the stochastic equations in \cite{Opanchuk12}, we apply
the same idea and project the noise due to non-linear losses over the
time-dependent condensate modes. Including $2-2$ and $1-2$ two-body
losses, plus one-body losses for the two states, we finally have for
the evolution during $dt$:
\begin{eqnarray}
  d\psi_1&=&-i dt \left[(\hat{h}_1- i K_1)+g_{11}|\psi_1|^2+(g_{12}-i K_{12})|\psi_2|^2 \right]\psi_1 + 
                             {\cal P}_{\phi_1}[{\rm \Delta_1}] \\
  d\psi_2&=&-i dt \left[(\hat{h}_2- i K_2)+(g_{22}- 2 i K_{22} ) |\psi_2|^2+ (g_{12} - i K_{12})|\psi_1|^2 \right]\psi_2 
                             \nonumber \\
         && + {\cal P}_{\phi_2}[{\rm \Delta_2}]\,, 
\end{eqnarray}
where $\hat{h}_j$ is the one-body Hamiltonian operator including the
kinetic energy and the external potential for the internal state $j$,
$g_{jk}=(4\pi \hbar^2 a_{jk})/m$ as above, $K_{12}=\gamma_{12}/2$,
$K_{22}=\gamma_{22}/4$ are two-body loss rate constants, and
$K_1=K_2=\tau^{-1}$ are one-body loss rate constants equal to the
inverse lifetime in the trap.  The projected noises have the
expressions
\begin{eqnarray}
  {\cal P}_{\phi_1}[{\rm \Delta_1}] &=& \phi_1(\mathbf{r},t) \left[ B_{12}(t) \sqrt{K_{12}I_{12}}+B_1(t)\sqrt{K_1}\right]\\
  {\cal P}_{\phi_2}[{\rm \Delta_2}] &=& \phi_2(\mathbf{r},t) \left[
                                        B_{22}(t)
                                        \sqrt{4K_{22}I_{22}}+B_{12}(t)
                                        \sqrt{K_{12}I_{12}}+B_2(t)\sqrt{K_2}\right]
\end{eqnarray}
where the $B_{12}(t)$, $B_{22}(t)$, $B_{1}(t)$ and $B_{2}(t)$ are
independent $\delta$-correlated complex Gaussian noises of variance
$dt$, e.g. $\langle B_{12}^\ast(t) B_{12}(t')\rangle=\delta(t-t')dt$,
and
$I_{jk}=\int \mathbf{dr} |\phi_j(\mathbf{r},t)
\psi_k(\mathbf{r},t)|^2$.
We have tested this method by comparing its results with an
exact solution of the two-mode model with losses \cite{Li08}. Details
can be found in \ref{sec:appendixSimul}.

\subsection{Simulation results}

\begin{figure}
\begin{center}
  \includegraphics[width=0.495\linewidth]{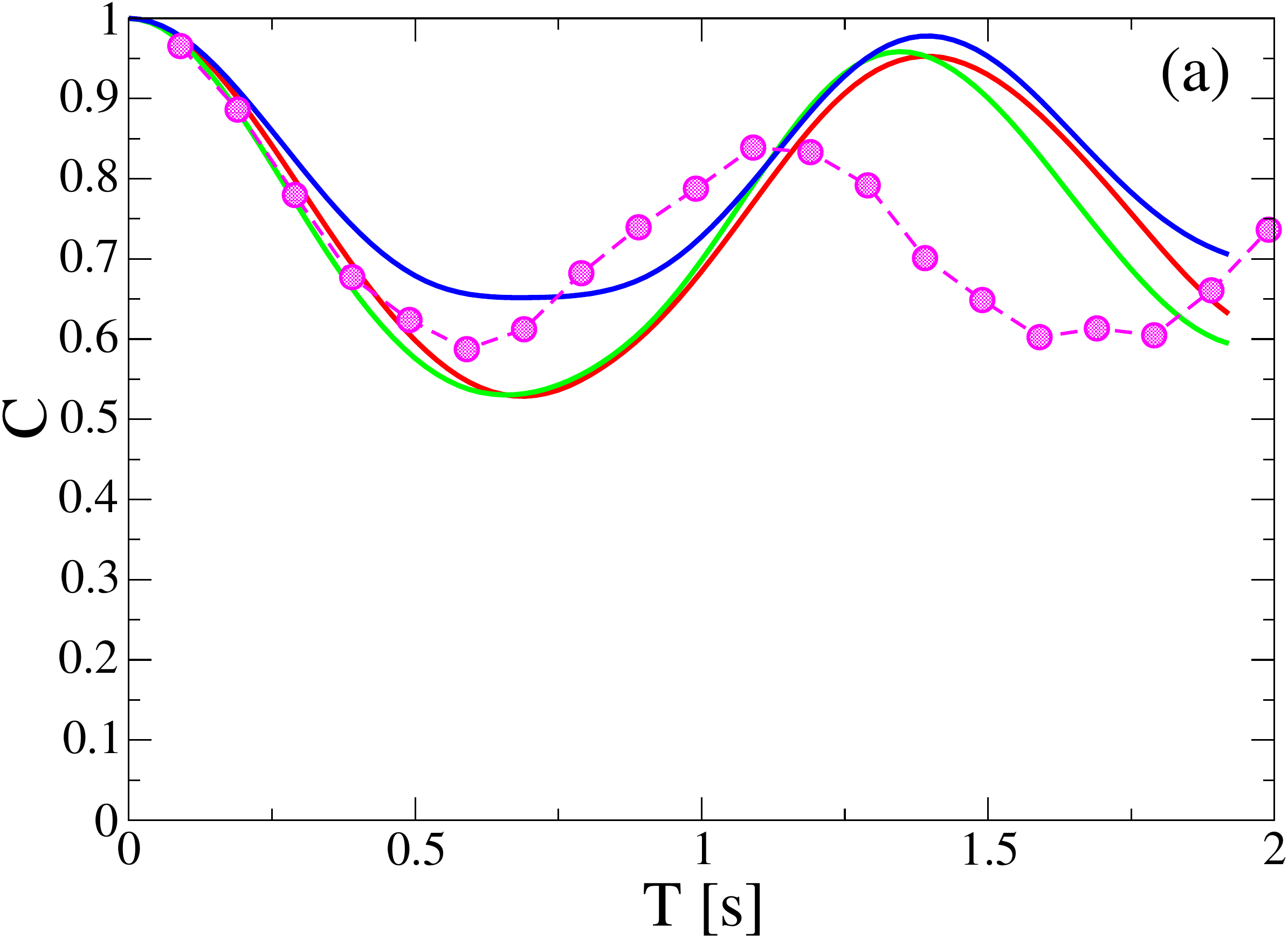}\includegraphics[width=0.505\linewidth]{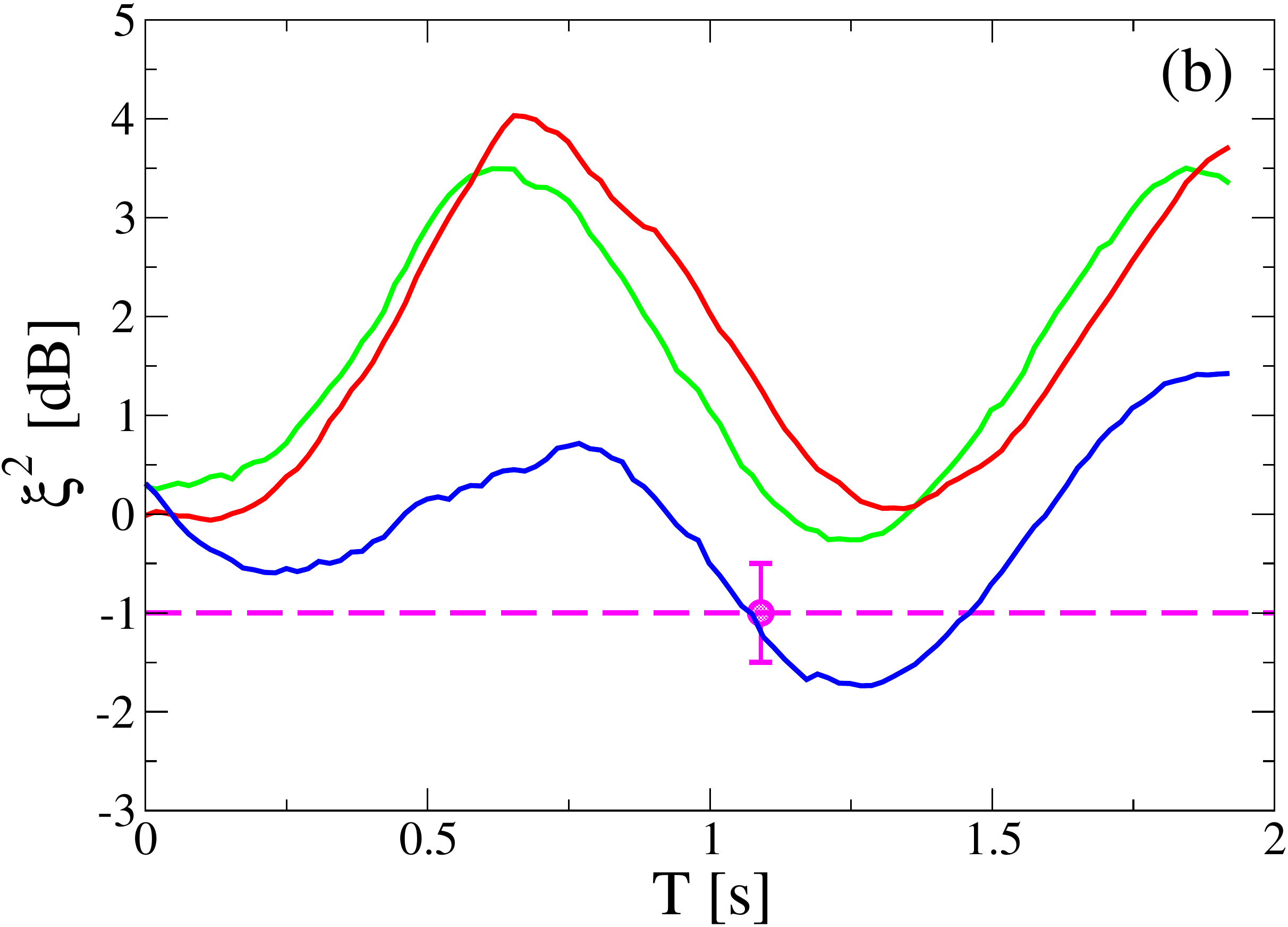}
  \caption{Contrast (a) and Spin squeezing (b) as a function of time.
    Solid lines: Wigner simulations, including spatial dynamics,
    quantum spin dynamics and particle losses, for three different
    choices of the scattering lengths.  Symbols: experiment (for (b),
    the experimental squeezing is measured at the time corresponding
    to the first contrast revival). The initial atom number is
    $N=10^4$. Trap frequencies
    $\omega_{x,y,z}=2\pi\times(2.9,92,74)$Hz.  Lifetime
    $\tau=5$s. Two-body loss rate constants
    $\gamma_{22}=8.1\times10^{-14}$cm${}^{3}/$s and
    $\gamma_{12}=1.51\times10^{-14}$cm${}^{3}/$s. Scattering lengths
    in Bohr radii units: $a_{11}=100.4$ for the three curves.  Red
    curve $a_{22}=95.44$ \cite{Egorov13}, $a_{12}=98.00$
    \cite{Egorov13}.  Green curve $a_{22}=95.00$ \cite{Mertes07},
    $a_{12}=97.66$ \cite{Mertes07}.  Blue curve $a_{22}=95.68$
    \cite{KokkelmansPC}, $a_{12}=97.66$ \cite{Mertes07}.  We used 800
    realizations for the Wigner simulation. The statistical
    uncertainty is around $10\%$ for the spin squeezing, corresponding
    to 0.4dB on the figure. The spatial grid had $128\times8\times8$
    points in the three directions and the initial temperature is
    zero. At the squeezing time in the simulation $T\simeq 1.37$s,
    approximately $\Nf \simeq 6000$ atoms are left in the trap, in a
    proportion $N_{1f}/N_{2f} \simeq 5/3$ for the two states.
\label{fig:Wigner}
}
\end{center}
\end{figure}

In Fig.~\ref{fig:Wigner} we show the results of the projected Wigner
simulation for an initial atom number $\Ni=10^4$ and three different
choices of the scattering lengths $a_{12}$ and $a_{22}$ chosen among
published values that differ by $0.7\%$ at most.
Fig.~\ref{fig:Wigner}(a) shows the contrast and
Fig.~\ref{fig:Wigner}(b) shows the squeezing as a function of
time. The experimental data for similar parameters is shown as symbols
for comparison.  Given that the demixing dynamics which induces the
squeezing is driven by the small differences between the scattering
lengths, it is not surprising that, in the absence of an external
state-dependent potential imposing the spatial separation
\cite{Hall98,Riedel10}, the squeezing result is very sensitive to the
precise values of the scattering lengths. To estimate the effective
nonlinearity for the different choices of the scattering lengths, we
calculated the parameter $\chi$ of the one-axis twisting Hamiltonian
at the stationary state in our geometry.  We obtain
$\chi=7.5\times10^{-5}$s${}^{-1}$ with the scattering length values
from \cite{Egorov13} (red curve), $\chi=7.3\times10^{-5}$s${}^{-1}$
with the values from \cite{Mertes07} (green curve), and
$\chi=24.6\times10^{-5}\,\mbox{s}^{-1}$ for the combination
\cite{Mertes07}-\cite{KokkelmansPC} (blue curve).

Comparing simulations and experiment, the contrast oscillations in the
experiment have a smaller amplitude and shorter period than in the
simulations (Fig.~\ref{fig:Wigner}(a)), as was already observed with the GPE
simulations. For the squeezing factor, the simulations do not allow a
quantitative comparison to experiment due to their strong
dependence on the scattering lengths. As shown in
Fig.~\ref{fig:Wigner}(b), depending on the choice of the scattering
length values, and despite their relatively high accuracy (compared to
the values available for other elements), the prediction at the
revival time varies between no squeezing at all and about -1.8\,dB.

We conclude that spontaneous squeezing in our geometry is compatible
with the results of our simulations although we cannot reproduce all
the features of the experimental data. If a quantitative agreement for
the contrast dynamics can be attained, it would be interesting to use
the extreme sensitivity to the scattering lengths to infer very
precise values for them from the experiment, similar to
\cite{Egorov13}.

\section{Conclusion and outlook}

Our results show that nonclassical spin dynamics occur spontaneously
in a two-component BEC, and can produce spin squeezing in BECs with
sizeable atom numbers. This supports the notion of squeezing as a
naturally occuring form of entanglement. In order to use this
squeezing as a resource for quantum metrology in particular, a higher
level of squeezing is desirable. Our results suggest several possible
routes.  The first is to accelerate the component separation, so that
squeezing would be produced on a faster timescale, before particle
losses become dominant. To do this, it would suffice to induce a small
asymmetry between the trapping potential of the two states at the
beginning of the sequence, in order to help the dynamics to start. It
has indeed been shown that this greatly enhances and accelerates the
spatial separation \cite{Ockeloen13}. In our case, this asymmetry
could come from the combination of the quadratic Zeeman effect with
gravity. This leads to a displacement of the center of the trapping
potentials for the two clock states that depends on the difference
between the field at the bottom of the trap and the magic field
\cite{Rosenbusch09}. Therefore, by scanning the magnetic field at the
trap bottom, one could displace the position of the two states and
study its influence on the spatial dynamics. In the same spirit, it
would be interesting to study whether the component separation can be
improved by modifying the aspect ratio of the trap, perhaps
dynamically.

Another path would be to act on the asymmetric two-body losses
themselves to reduce their rate. A possible approach could be to use
microwave dressing during the interrogation time to shift the
$\ket{2,2}$ state upward and induce an energy difference between the
transitions $\ket{2,1}\rightarrow\ket{2,2}$ and
$\ket{2,1}\rightarrow\ket{2,0}$, thereby reducing the two-body
collision rate in state $\ket{2,1}$. This could be accomplished using
a one-photon dressing with a $\sigma$-polarized microwave field. Of
course, one would need to check that this additional coupling does not
introduce too much noise on the clock transition. Observing a
reduction of these losses would also be an interesting subject in itself.

\ack We thank Markus Oberthaler, Philipp Treutlein and Christian Gross
for inspiring discussions.  This work was supported by the
D\'el\'egation G\'en\'erale de l'Armement (DGA) through the ANR ASTRID
program (contract no. ANR-14-ASTR-0010, project ``eeTACC''), the
European Research Council (ERC), (GA 671133, Advance Grant
``EQUEMI''), and the Institut Francilien pour la Recherche sur les
Atomes Froids (IFRAF).

\appendix
\section{Imaging and Calibration}
\label{sec:appendixExp}

\paragraph{\textbf{Detection}\\}

\begin{figure*}[tb!]
\centering
\includegraphics[width=0.47\columnwidth]{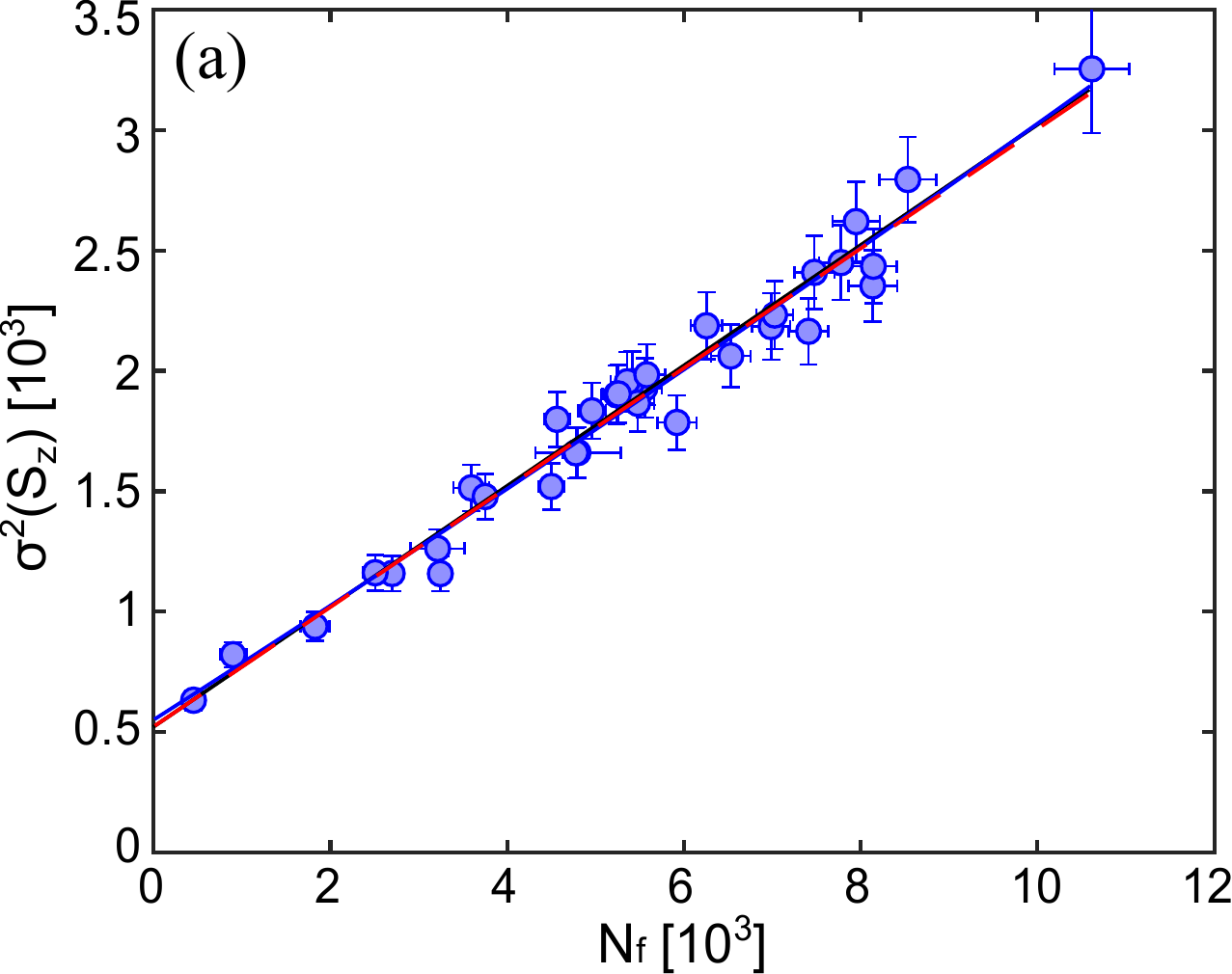}
\includegraphics[width=0.47\columnwidth]{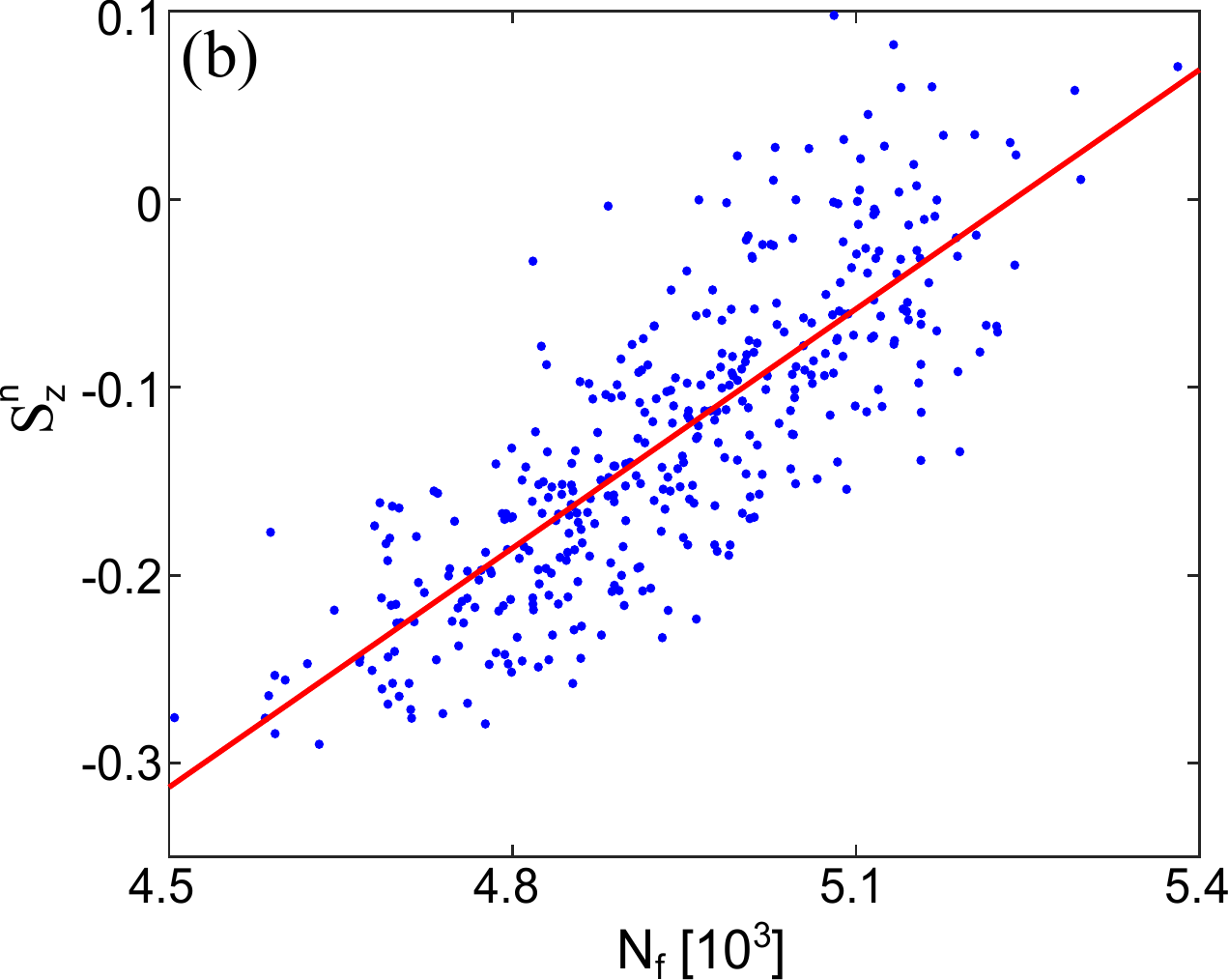}
\caption{(a) Validation of the detection calibration. The variance of
  $S_z$ measured right after a single resonant $\frac{\pi}{2}$-pulse
  is plotted as a function of the detected atom number. Both the
  parabolic (blue line) and linear (red dashed line) fits give a
  linear part compatible with the standard quantum limit and a
  detection noise $\sigma_{det}\approx33$ atoms. (b) Atom number
  correlation for $\alpha = 90^\circ$.}
\label{fig:detection}
\end{figure*}

For accurate, low-noise atom number measurement, we use saturated
absorption imaging \cite{Reinaudi07} along the slow $x$ axis, combined
with spatially separated detection of both clock states in the same
image, and employ back-illuminated deep depletion CCD camera with
$>90\%$ quantum efficiency (Andor iKon M 934-BRDD).  After the
sequence but still in trap, atoms in $\ket{1}$ are adiabatically
transferred to the untrapped $\ket{F = 2, m_F =0}$ state in 2\,ms by a
strong MW pulse while the atomic resonance is swept by ramping the
magnetic bias field $B_x$.  Adiabaticity is ensured by Blackman pulse
shape for the MW power and a half-Blackman ramp for $B_x$ ($\pm50$mG
around the pseudo-magic field).  During this pulse, atoms initially in
state $\ket{1}$ start to fall under the action of gravity. 50\,$\mu$s
later, the trapping magnetic fields are turned off in order to release
the remaining atoms, such that the two states can be spatially
discriminated after 23\,ms time of flight and imaged with a single 20
$\mu$s detection pulse. This way, frequency and power fluctuations of
the probe laser are in common mode for the two states, reducing
fluctuations in the detected population difference. Additionally, a
numerical frame re-composition algorithm is used to reduce optical
fringes \cite{Ockeloen10}.  The column density is then derived taking
into account the high saturation correction \cite{Reinaudi07}. With
this imaging procedure, the background noise of our imaging system is
about 33 atoms for each of the spin components (cf.\
Fig.~\ref{fig:detection} (a)).

Great care is taken to calibrate the detection system.  The
calibration method is similar to \cite{Riedel10}, and consists in
comparing the variance of $S_z$ for an ensemble of $N$ uncorrelated
atoms in a coherent superposition with the standard quantum limit that
scales as $\frac{N}{4}$. To perform the calibration, $S_z$ is measured
directly after a single resonant $\frac{\pi}{2}$-pulse, and the
measured variance is plotted as a function of the total detected atom
number $\Nf$. The data is fitted with
$\sigma^2(S_z) = \sigma_{det}^2+\sigma _{qpn}^2 \overline{N} +
\sigma_{tech}^2 \overline{N}^2$, where $\sigma_{det}$ is the detection
noise, $\sigma_{qpn}$ represents the quantum projection noise and
$\sigma_{tech}$ accounts for the possible preparation noise.  As shown
in Fig.~\ref{fig:detection} (a), the atomic noise exhibits the
expected linear behavior with a slope of 
$\sigma _{qpn}^2=0.248\pm 0.03$ and a neglegible quadratic component,
  $\sigma_{tech}^2=(1\pm 3)\dip{-6}$, confirming that our detection
is projection noise limited.

\paragraph{\textbf{Data analysis}\\}
In order to link the measured atom numbers to the squeezing factor
and the spin noise distribution, the following data analysis is
performed for each tomography angle $\alpha$. The notation
$\overline{X}$ corresponds to the average of the fluctuating quantity
X, and the error bars correspond to one standard deviation.

\begin{itemize}
\item Shots whose total atom numbers $\Nf = N_1 + N_2$ differ from
  their mean $\overline{N}$ by more than 3 standard deviations are
  discarded. These outliers can be due to problems during the image
  acquisition or the laser locks, and in practice this concerns about
  1$\%$ of the data.

\item The normalized population difference
  $S_z^n=\frac{N_2-N_1}{2\Nf}$ and the angle between the collective
  spin and the equator of the Bloch sphere
  $\theta_c=\text{asin}(\frac{S_z^n}{C/2})$ are derived, where $C$ is
  the contrast of the Ramsey interferometer, which has been
    measured separately. The population difference is normalized in
  order to reduce its dependency to shot-to-shot total atom number
  fluctuations.

\item A correlation between population difference and total atom
  number exists because of the atom number dependency of the atomic
  frequency ($1.5\times10^{-4}$ Hz per atom) via the collisional shift
  \cite{Harber02}, and can be estimated by fitting the distribution
  $S_z^n$ vs $\Nf$ (cf.\ Fig.~\ref{fig:detection} (b)). Since we can
  measure this correlation for each shot, we can legitimately correct
  the data accordingly. Namely,
  \begin{equation}
    S_{z,corr}^n(i) = S_z^n(i) - s_p\times (\Nf(i)-\overline{\Nf}),
    \label{eq:correction}
  \end{equation}
  where $s_p=\frac{dS_z^n}{d\Nf}$ is the slope measured on
  Fig.~\ref{fig:detection} (b) and $i$ represents one of the 300 shots
  at a given rotation angle. This slope, and thus the correction,
  depends on the analysis pulse duration and happens to vanish around
  $\alpha = 2.5^\circ$; it therefore does no affect the squeezing
  factor.

\item The variance of the normalized population difference times the
  mean atom number $\overline{\Nf}\sigma^2(S_{z,corr}^{n})$ is
  derived. At this point, we also check that the Allan variance of
  $S_{z,corr}^{n}$ integrates as white frequency noise in order to be
  sure that there is no drift that could worsen the results.

\item The detection noise estimated in Fig.~\ref{fig:detection} (a) is
  removed from the data:
  \begin{equation}
    \Delta S_z^2 = \overline{\Nf}^2 \sigma^2(S_{z,corr}^{n}) - \sigma_{det}^2(S_z)
  \end{equation}

\item Finally, the fact that the collective spin ends up below the
  equator of the Bloch sphere leads to an underestimation of the spin
  noise by a factor $\text{cos}(\overline{\theta_c})^2$ which has to
  be taken into account as explained in
  sec.~\ref{sec:contrast}. $\Delta S_z^2$ is then normalized by the
  quantum projection noise $\Nf/4$, leading to the final number
  squeezing and squeezing factor \cite{Wineland94}
  \begin{equation}
    \mathcal{V}^2 = \frac{4\Delta S_z^2}{\overline{\Nf}\text{cos}(\overline{\theta_c})^2} \text{ and } \xi^2=\frac{4\Delta S_z^2}{\overline{\Nf}C^2\text{cos}(\overline{\theta_c})^2}.
    \label{eq:variance_norm}
  \end{equation}
\end{itemize}

\paragraph{\textbf{Analysis pulse calibration}\\}
In order to perform the state tomography, one could in principle apply
the correction pulse discussed in sec.~\ref{sec:contrast} prior to the
analysis pulse, which would simply need to be phase shifted with
respect to the preparation pulse. However, because of microwave
inhomogeneity and position fluctuation of the trapped BEC, this
correction pulse would introduce additional noise in the spin
tomography sequence. The idea is then to keep the Bloch vector below
the equator of the Bloch sphere, and align the Rabi vector of the
analysis pulse with the Bloch vector to rotate the noise distribution
about its center. The z-component of the Rabi vector is given by the
detuning between the considered pulse $\nu_{lo}$ and the instantaneous
atomic frequency $\nu_{at}$ at the time at which the pulse is
applied. The azimuthal angle is controlled via the phase shift with
respect to the first pulse. First one needs to measure $\nu_{at}$,
which is the frequency for which the resulting transition probability
does not depend on the local oscillator phase-shift when the analysis
pulse is a $\pi-$pulse. The detuning $\nu_{lo}-\nu_{at}$ is then
simply given by $\pm \frac{\Omega_R}{2\pi} \text{tan}(\theta_c)$. The
$"\pm"$ sign gives the direction of rotation performed during the
tomography. Finally, the phase-shift that aligns the two vectors is
the one that makes the resulting transition probability independent of
the analysis pulse duration for the previously derived frequency.

\section{Test of the projected Wigner method}
\label{sec:appendixSimul}
\begin{figure}
\begin{center}
 \includegraphics[width=0.5\linewidth]{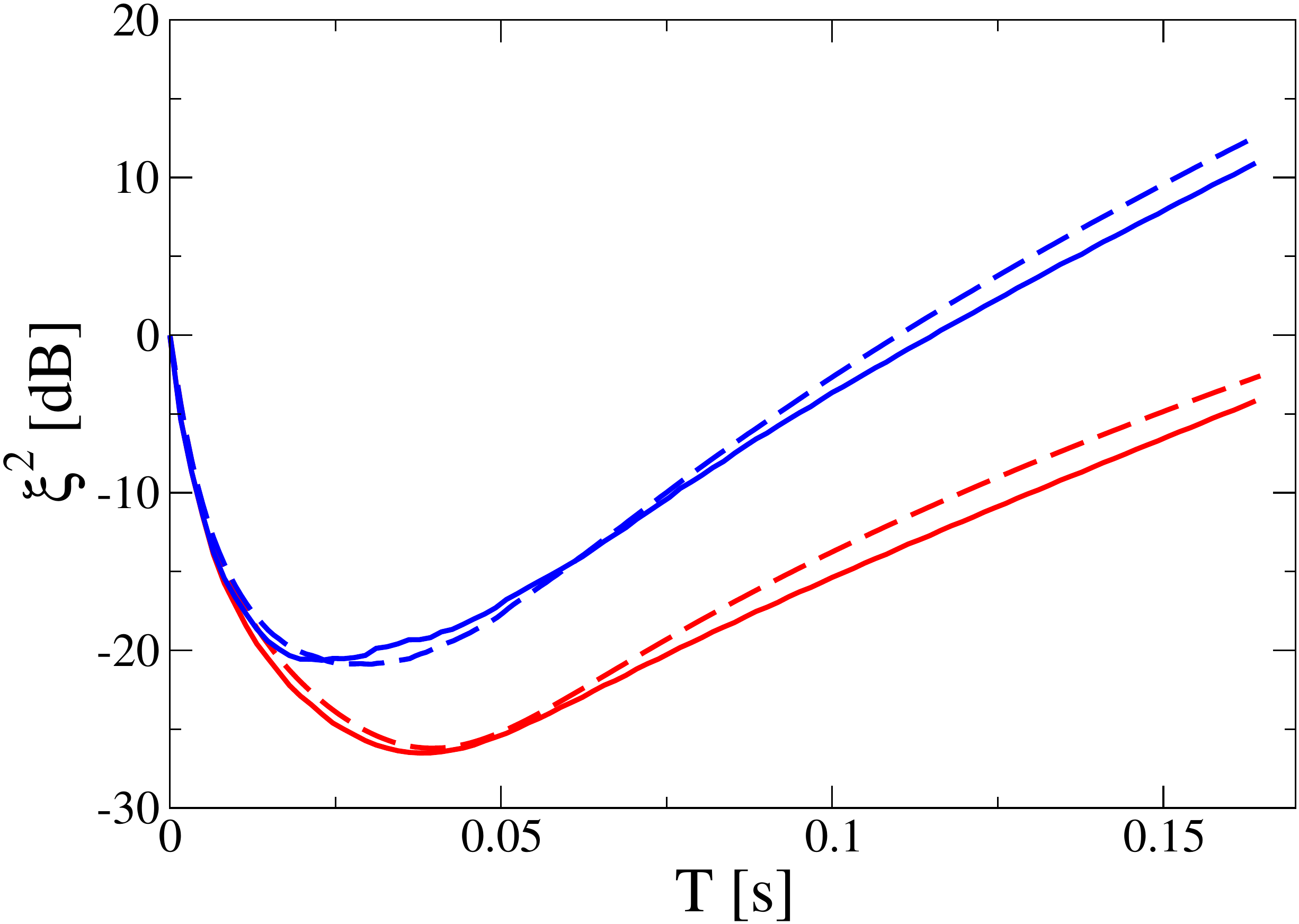} 
 \caption{Test of the projected Wigner method against an exact
   solution of the two-mode model with and without losses, in a
   situation in which the spatial dynamics of the condensate wave
   functions is not excited. Spin squeezing as a function of time
   without losses (lower curves) and with asymmetric two-body
   losses (upper curves).  Comparison between the projected Wigner
   simulation (solid lines) and a two-mode Monte Carlo simulation
   (dashed lines) with 1600 realizations.  The initial atom number is
   $N=10^{4}$. Trap frequencies
   $\omega_{x,y,z}=2\pi\times(2.9,92,74)$Hz. Lifetime
   $\tau=5$s. Two-body loss rate constants
   $\gamma_{22}=8.1\times10^{-14}$cm${}^{3}/$s and
   $\gamma_{12}=1.51\times10^{-14}$cm${}^{3}/$s. Scattering lengths
   $a_{11}(0^-)=100.4 a_0$, $a_{11}(0^+)=a_{22}(0^+)=2a_{11}(0^-)$ and
   $a_{12}=0$. We used 400 realizations for the Wigner simulation. At
   the end of the simulation, $\omega_x T=3$, approximately
   $\Nf \simeq 9400$ atoms are left in the trap, in a proportion
   $N_{1f}/N_{2f} \simeq 0.515/0.485$ for the two states. For the
   two-mode model we used $\chi=0.064$s${}^{-1}$, calculated at steady
   state after the pulse.
   \label{fig:Test2M}}
\end{center}
\end{figure}
To test the projected Wigner method, we compared its results with an
exact solution of the two-mode model with losses \cite{Li08}, in a
situation in which the spatial dynamics of the condensate wave
functions is not excited. To this end we choose $g_{ab}=0$,
$g_{aa}=g_{bb}=g$, and the value of $g$ was doubled after the mixing
pulse to keep the mean field constant.  For the two-mode model we
calculated the corresponding parameter $\chi$ of the one-axis twisting
Hamiltonian and solved the corresponding master equation using the
Monte Carlo wave function method with a large number of realizations.
As $g_{ab}=0$, the nonlinearity is large and the squeezing generation
is fast with respect to the losses, so that the squeezing factor gets
very small which makes a good test for the projected Wigner method. As
shown in Fig.~\ref{fig:Test2M}, we find a good agreement between the
two methods.

\section*{References}

\bibliographystyle{njpph2}
\bibliography{lascool,local}

\begin{thebibliography}{10}

\bibitem{Kawaguchi12}
Kawaguchi Y and Ueda M 2012 {\it Physics Reports\/} {\bf 520} 253

\bibitem{Esteve08}
Esteve J, Gross C, Weller A, Giovanazzi S and Oberthaler M~K 2008 {\it
  Nature\/} {\bf 455} 1216

\bibitem{Gross10}
Gross C, Zibold T, Nicklas E, Est{\`e}ve J and Oberthaler M~K 2010 {\it
  Nature\/} {\bf 464} 1165

\bibitem{Riedel10}
Riedel M~F, B{\"o}hi P, Li Y, H{\"a}nsch T~W, Sinatra A and Treutlein P 2010
  {\it Nature\/} {\bf 464} 1170

\bibitem{Luecke11}
L{\"u}cke B, Scherer M, Kruse J, Pezz{\'e} L, Deuretzbacher F, Hyllus P, Topic
  O, Peise J, Ertmer W, Arlt J, Santos L, Smerzi A and Klempt C 2011 {\it
  Science\/} {\bf 334} 773

\bibitem{Schmied16}
Schmied R, Bancal J~D, Allard B, Fadel M, Scarani V, Treutlein P and Sangouard
  N 2016 {\it Science\/} {\bf 352} 441

\bibitem{Kitagawa93}
Kitagawa M and Ueda M 1993 {\it Phys.~Rev.~A\/} {\bf 47} 5138

\bibitem{Wineland94}
Wineland D~J, Bollinger J~J, Itano W~M and Heinzen D~J 1994 {\it Phys. Rev.
  A\/} {\bf 50} 67

\bibitem{Gross12}
Gross C 2012 {\it J.~Phys.~B: At.\ Mol.\ Opt.\ Phys.\/} {\bf 45} 103001

\bibitem{Abend16}
Abend S, Gebbe M, Gersemann M, Ahlers H, M\"untinga H, Giese E, Gaaloul N,
  Schubert C, L\"ammerzahl C, Ertmer W, Schleich W~P and Rasel E~M 2016 {\it
  Phys. Rev. Lett.\/} {\bf 117} 203003

\bibitem{Hardman16}
Hardman K~S, Everitt P~J, McDonald G~D, Manju P, Wigley P~B, Sooriyabandara
  M~A, Kuhn C~C~N, Debs J~E, Close J~D and Robins N~P 2016 {\it
  Phys.~Rev.~Lett.\/} {\bf 117} 138501

\bibitem{Sorensen01}
S{\o}rensen A, Duan L~M, Cirac J~I and Zoller P 2001 {\it Nature\/} {\bf 409}
  63

\bibitem{Haine14}
Haine S~A, Lau J, Anderson R~P and Johnsson M~T 2014 {\it Phys.~Rev.~A\/} {\bf
  90} 023613

\bibitem{Pezze16}
Pezz{\`e} L, Smerzi A, Oberthaler M, Schmied R and Treutlein P 2017 {\it
  arxiv:1609.01609\/}

\bibitem{Li09}
Li Y, Treutlein P, Reichel J and Sinatra A 2009 {\it Eur.~Phys.~J.~B\/} {\bf
  68} 365

\bibitem{Sinatra12a}
Sinatra A, Dornstetter J~C and Castin Y 2012 {\it Front. Phys.\/} {\bf 7} 86

\bibitem{Maussang10}
Maussang K, Marti G~E, Schneider T, Treutlein P, Li Y, Sinatra A, Long R,
  Est\`eve J and Reichel J 2010 {\it Phys. Rev. Lett.\/} {\bf 105} 080403

\bibitem{Mertes07}
Mertes K, Merrill J, Carretero-Gonz\'alez R, Frantzeskakis D, Kevrekidis P and
  Hall D 2007 {\it Phys.~Rev.~Lett.\/} {\bf 99} 190402

\bibitem{Harber02}
Harber D~M, Lewandowski H~J, McGuirk J~M and Cornell E~A 2002 {\it
  Phys.~Rev.~A\/} {\bf 66} 053616

\bibitem{Szmuk15}
Szmuk R, Dugrain V, Maineult W, Reichel J and Rosenbusch P 2015 {\it
  Phys.~Rev.~A\/} {\bf 92} 012106

\bibitem{Hall98}
Hall D~S, Matthews M~R, Ensher J~R, Wieman C~E and Cornell E~A 1998 {\it
  Phys.~Rev.~Lett.\/} {\bf 81} 1539

\bibitem{Egorov11}
Egorov M, Anderson R~P, Ivannikov V, Opanchuk B, Drummond P, Hall B~V and
  Sidorov A~I 2011 {\it Phys. Rev. A\/} {\bf 84} 021605

\bibitem{Nicklas15}
Nicklas E, Muessel W, Strobel H, Kevrekidis P and Oberthaler M 2015 {\it
  Phys.~Rev.~A\/} {\bf 92} 053614

\bibitem{Lacroute10}
Lacroute C, Reinhard F, {Ramirez-Martinez} F, Deutsch C, Schneider T, Reichel J
  and Rosenbusch P 2010 {\it {IEEE} Transactions on Ultrasonics, Ferroelectrics
  and Frequency Control\/} {\bf 57} 106

\bibitem{Deutsch10}
Deutsch C, Ramirez-Martinez F, Lacro\^ute C, Reinhard F, Schneider T, Fuchs
  J~N, Pi\'echon F, Lalo\"e F, Reichel J and Rosenbusch P 2010 {\it
  Phys.~Rev.~Lett.\/} {\bf 105} 020401

\bibitem{Treutlein04}
Treutlein P, Hommelhoff P, Steinmetz T, H{\"a}nsch T~W and Reichel J 2004 {\it
  Phys.~Rev.~Lett.\/} {\bf 92} 203005

\bibitem{Ramirez10}
{Ramirez-Martinez} F, Lours M, Rosenbusch P, Reinhard F and Reichel J 2010 {\it
  {IEEE} Transactions on Ultrasonics, Ferroelectrics and Frequency Control\/}
  {\bf 57} 88

\bibitem{Papp08}
Papp S~B, Pino J~M and Wieman C~E 2008 {\it Phys.~Rev.~Lett.\/} {\bf 101}
  040402

\bibitem{Tojo10}
Tojo S, Taguchi Y, Masuyama Y, Hayashi T, Saito H and Hirano T 2010 {\it
  Phys.~Rev.~A\/} {\bf 82} 033609

\bibitem{Egorov13}
Egorov M, Opanchuk B, Drummond P, Hall B, Hannaford P and Sidorov A 2013 {\it
  Phys.~Rev.~A\/} {\bf 87} 053614

\bibitem{Reinaudi07}
Reinaudi G, Lahaye T, Wang Z and Gu{\'e}ry-Odelin D 2007 {\it Opt. Lett.\/}
  {\bf 32} 3143

\bibitem{Li08}
Li Y, Castin Y and Sinatra A 2008 {\it Phys.~Rev.~Lett.\/} {\bf 100} 210401

\bibitem{Opanchuk12}
Opanchuk B, Egorov M, Hoffmann S, Sidorov A~I and Drummond P~D 2012 {\it
  Europhys.~Lett.\/} {\bf 97} 50003

\bibitem{Sinatra02}
Sinatra A, Lobo C and Castin Y 2002 {\it J.~Phys.~B: At.\ Mol.\ Opt.\ Phys.\/}
  {\bf 35} 3599

\bibitem{KokkelmansPC}
Kokkelmans S~J~J~M~F 2013 {\it private communication, cited in
  \cite{Egorov13}\/}

\bibitem{Ockeloen13}
Ockeloen C~F, Schmied R, Riedel M~F and Treutlein P 2013 {\it
  Phys.~Rev.~Lett.\/} {\bf 111} 143001

\bibitem{Rosenbusch09}
Rosenbusch P 2009 {\it Applied Physics B: Lasers and Optics\/} {\bf 95} 227

\bibitem{Ockeloen10}
Ockeloen C~F, Tauschinsky A~F, Spreeuw R~J~C and Whitlock S 2010 {\it
  Phys.~Rev.~A\/} {\bf 82} 061606

\end{thebibliography}
\end{document}